\documentclass[aps,pra,twocolumn,showpacs,superscriptaddress,groupedaddress]{revtex4-2}

\usepackage{amsmath,braket,amssymb,natbib,graphicx,float,amsthm,xcolor,outlines,mathrsfs,multirow,hhline}
\usepackage{multirow}
\usepackage{mathtools}
\usepackage{bm}

\usepackage[english]{babel}
\usepackage[utf8x]{inputenc}
\usepackage[T1]{fontenc}
\usepackage{listings}
\usepackage[colorinlistoftodos]{todonotes}
\usepackage[colorlinks=true, allcolors=blue]{hyperref}
\begin{document}
\title{Dynamics of a multipartite hybrid quantum system with  beamsplitter, dipole-dipole, and Ising interactions}

\author{Pradip Laha}
\email{plaha@uni-mainz.de}
\affiliation{Department of Physics, IIT Madras, Chennai 600036, India \\
                Institute of Physics, Johannes-Gutenberg University of Mainz, Staudingerweg 7, 55128 Mainz, Germany}


\begin{abstract}
The possibility  of exploiting heterogeneous quantum systems to high precision,  for storing, processing, and transmitting  information, makes them ideal candidates for multi-tasking purposes in quantum communication. Appropriate quantum systems involving a judicious choice of interactions which augment each other, are potentially useful for probing deep into quantum regimes.  Here, we make use of one such hybrid bipartite quantum model, with  one subsystem made of a pair of qubits and another comprising  a pair of  oscillators,  to study the entanglement dynamics,  and the entanglement  transfer between discrete  and continuous variables. Our basic model is the standard double Jaynes-Cummings  system, which is known to support both entanglement transfer and entanglement sudden death, under suitable conditions.  In this work, we generalise this model to include further experimentally relevant interactions, such as the beamsplitter-type exchange interaction between the oscillators, and dipole-dipole and Ising-type interactions between the qubits. The manner in which various interactions and  initial  oscillator states  affect the entanglement dynamics, is examined theoretically,  for generic experimental conditions. Using exact analytical solutions, we show that compared to   the beamsplitter or dipole-dipole interaction,  the Ising interaction can have a significant positive impact on  entanglement sudden death and birth,  and postponement of the onset of these phenomena, apart from  producing substantial reduction in the time duration of the death.
\end{abstract}

\maketitle


\section{\label{sec:intro} Introduction}
A wide range of quantum mechanical tasks, such as simulation of complex quantum dynamics~\cite{buluta_science_2009,cirac_goals_2012,blatt_quantum_2012,houck_chip_2012,lamata_2018,wilkinson_superconducting_2020}, processing of quantum information~\cite{nielsen},  creating secure communication~\cite{gisin_RMP_2002,Sergienko} and high-precision  sensing~\cite{schmidt_science_2005,giovannetti_advances_2011,hempel_entanglement_2013,ockeloen_PRL_2013,wolfgramm_entanglement-enhanced_2013,degen_RMP_2017}, can be  carried out efficiently using  platforms that combine both continuous variable (CV) and discrete variable (DV) quantum systems.  These could include interacting subsystems such as those involving combinations of  photons, acoustic or optical phonons, natural and artificial atoms with semiconductor quantum dots, ions,  electronic and nuclear spins, polaritons, magnons,  mesoscopic superconducting or transmon qubits, etc.~\cite{Kjaergaard_2020,conner_APL_2021,koch_PRA_2007,oliver_welander_2013,place_new_2021,kurizki_quantum_2015,clerk_hybrid_2020,lambert_AQT_2020,blais_RMP_2021}.  Different hybrid platforms are identified for performing specific quantum tasks with the  desired precision.  For instance, fiber optics is  ideal for long distance quantum communication, weakly interacting electronic or nuclear spins are excellent for devising  long-lived quantum memories due to their long coherence times, while macroscopic Josephson-junction-based superconducting circuits, owing to their strong, controllable couplings with microwave photons,  are ideal for rapidly processing quantum information. Hybrid quantum systems have therefore emerged as ideal candidates  for performing multiple quantum tasks simultaneously and with considerable control, providing optimal platforms that circumvent the  deficiencies of  individual CV or DV platforms, at the same time enhancing their individual efficiencies~\cite{xiang_RMP_2013,kurizki_quantum_2015,Cottet_2017,blais_RMP_2021,clerk_hybrid_2020,lambert_AQT_2020,li_PRL_2020}. The major hurdle  for developing such heterogeneous quantum systems is that the natural excitation frequencies of the degrees of freedom of seemingly disparate platforms are very different, thus making it extremely hard to generate strong couplings between them. Hence, much of the effort in this rapidly growing field is devoted towards finding novel  ways to create stronger couplings.  

 The most common interaction in platforms involving CV systems is the linear beamsplitter-type (BS) exchange interaction. For two  CV systems A and B, this interaction is given by the Hamiltonian (setting $\hbar = 1$)
\begin{align}
 \displaystyle H_{\text{BS}} =  g_{\text{b}} \left(a \, b^{\dagger} + a^{\dagger}\, b\right),
\label{eqn:ham_bs}
\end{align}
where $a^\dagger\, (a$) and $b^{\dagger}\,(b$) are the creation/excitation (annihilation/de-excitation) operators for systems A and B, respectively. $g_{\text{b}}$ is the coupling strength between the two CV systems.  The unitary dynamics governed by $H_\text{BS}$, results in Rabi flops between modes A and B with frequency $\frac{g_{\text{b}}}{\pi}$~\cite{clerk_hybrid_2020}.
The BS interaction arises in a variety of realistic hybrid quantum systems.   For instance, in systems based on magnonics, 
it is used in the rotating wave approximation, to model the  interaction of magnetostatic modes with the microwave-frequency cavity modes~\cite{Lachance_Quirion_2019}. 
Yet again the BS term appears in circuit-QED hybrid Hamiltonians, where the two CV systems A and B correspond to the superconducting microwave photons and optical photons~\cite{lambert_AQT_2020}.

On the other hand, for DV systems (such as spins with states $\ket{\uparrow}$ and $\ket{\downarrow}$, qubits with states $\ket{0}$ and $\ket{1}$, or equivalently atoms with two effective energy levels $\ket{g}$ and $\ket{e}$, respectively), the  interactions among the  subsystems can be modelled either as the dipole-dipole exchange interaction or  the Ising-type interaction. For two such DV systems 1 and 2 these interactions are given by the Hamiltonians
\begin{align}
 \label{eqn:ham_qq}
  H_{\text{DD}} &= g_{\text{d}} \left(\sigma_{+}^{(1)} \sigma_{-}^{(2)} + \sigma_{-}^{(1)} \sigma_{+}^{(2)}\right), \\
  H_{\text{IS}} &= g_{\text{i}} \, \sigma_{z}^{(1)} \sigma_{z}^{(2)},
 \label{eqn:ham_is}
\end{align}
respectively. Here, for two-level atoms,  $\sigma_{+}^{(i)} = \ket{e_i}\bra{g_i}$, $\sigma_{-}^{(i)} = \ket{g_i}\bra{e_i}$ and $\sigma_{z}^{(i)} = \ket{e_i}\bra{e_i} - \ket{g_i}\bra{g_i}$ ($i=1,\,2$), and analogous interpretation of the Pauli matrices for  spin states and generic qubit states is apparent. $g_{\text{d}}$ and $g_{\text{i}}$ are, respectively, the coupling strengths of the dipole-dipole and Ising interactions.
A wide range of experimentally realisable multipartite systems have interactions  governed by these Hamiltonians. For instance, 
laser-trapped circular Rydberg atoms used in quantum simulators, make use of the strong interactions between the atoms to simulate an XXZ spin chain Hamiltonian~\cite{nguyen_PRX_2018}. The spin-spin interactions considered in this are the strong dipole-dipole and the Ising interactions. A proposal for a hybrid scheme in circuit QED, realized by integrating Rydberg atoms into superconducting circuits, also exploits the versatility of these  forms of interaction. 

Finally, the coupling between the CV and DV systems can be described by the well-known Jaynes-Cummings (JC) interactions.  In this work, we are interested in heterogeneous quantum systems involving two CV  and two DV systems. For  ease of nomenclature, we refer to the two DV systems as qubit 1 and qubit 2 respectively,  while the two CV oscillators  are  denoted by  A and B respectively. The inherently nonlinear JC-type interaction between qubit 1 and oscillator A is given by the well-studied  Hamiltonian
\begin{align}
 \label{eqn:ham_jc1}
 \displaystyle H^{(1)}_{\text{JC}} &= g_{\text{JC}}^{(1)} \left(a \,\sigma_{+}^{(1)} + a^\dagger \,\sigma_{-}^{(1)}  \right), 
\end{align}
where $g_{\text{JC}}^{(1)}$ is the JC interaction strength. Similarly, qubit 2 and oscillator B interact through another JC-type interaction Hamiltonian $H^{(2)}_{\text{JC}}$ with interaction strength $g_{\text{JC}}^{(2)}$.
The weak coupling   between natural atoms in cavity QED is overcome, for instance, by using  hybrid platforms in which superconducting qubits are embedded in a superconducting microwave resonator. 

The implementation of the many quantum tasks mentioned above relies heavily on the nonclassical properties exhibited by the quantum systems~\cite{nelsen_chuang}, primarily quantum entanglement~\cite{schrodinger,horodecki,GUHNE20091} which plays a vital role in several quantum information protocols in  cryptography,  teleportation,  key distributions, and computation  and simulation~\cite{ekart,riebe_teleportation,bennett_qkd,long_PRA_2002,bennett,deng_PRA_2003}. Although, in many quantum systems entanglement can be generated quite straightforwardly, the  challenge is to sustain it without significant decoherence (through spontaneous emission, dephasing, etc.) over a time interval which is long enough for it to be used as a genuine resource (see, for instance, \cite{zurekwh,maximilian}).  In order for the quantum entanglement to sustain against  such asymptotic losses, several schemes have been proposed in the literature. These include quantum error correction procedures  dedicated to detecting and eliminating specific errors that occur during the storage and processing of quantum information~\cite{suter_RMP_2016}, and encoding quantum information in a `decoherence-free subspace', a scheme that has already been implemented in  systems such as trapped ions, neutral atoms, charged quadrupole qubits,  superconducting devices, etc.~\cite{lidar_PRL_1998,monz_PRL_2009,lidar_book_2014,friesen_nature_2017}. Also, various schemes of entanglement distillation, state purification and concentration have been proposed theoretically, and demonstrated  experimentally (see, for instance~\cite{kwiat,pan1}).
Modern advances in  technology have further resulted in  significant increase in the coherence time  by efficiently isolating the quantum system from its environment. 

In complete contrast to such exponential decay of entanglement, there can also arise sudden death of entanglement  between two subsystems of a well-isolated  multipartite quantum system, where the entanglement is lost completely in a finite time. This is more destructive than generic asymptotic decay. It was shown that the entanglement between two initially entangled atoms with no direct interaction in the governing Hamiltonian, placed in two separate cavities can exhibit entanglement sudden death
(ESD)~\cite{eberly_prl,Y_na__2006,Eberly_PRL_2006,eberly_oc,Y_na__2007,eberly2,eberly_qic,eberly1}, a phenomenon that has already been observed in experiments  involving entangled photon pairs~\cite{almeida,Salles_PRA_2008,xu_esd,Pramanik_PRA_2019}, atomic ensembles~\cite{laurat}, trapped ions~\cite{Barreiro_2010}, and solid state systems~\cite{Wang_PRB_2018}. An allied topic is entanglement transfer between subsystems, as for instance, between the qubit-qubit subsystem and the oscillator-oscillator subsystem~\cite{Lopez_PRL_2008}.  Since its first appearance, ESD and entanglement transfer has been studied extensively in  the literature on a wide range of quantum systems (such as cavity and circuit QED, all-optical setup using NOT operations, trapped ions, etc.) involving qubit pairs, qutrit pairs, qubit-qutrit pairs, finite spin chains, multipartite systems of qubits interacting with bosonic fields or two identical fermions interacting with a global bosonic environment or open system dynamics~\cite{Lopez_PRL_2008,Rau_EPL_2008,ali_manipulating_2008,Man_2008,carvalho,Yamamoto_PRA_2008,Qiang_2008,Chan_2009,Das_PRA_2010,MIH_2011,Kim_Nat_Phys_2012,Singh_JOSAB_17,deng_sci_rep_2017,chathavalappil_schemes_2019,Sadiek_OE_2019,hang_entanglement_2020,YMeng_PRA_2020,Bussandri_2020,Ghosal_PRA_2020,Xue_Ent_2020,sharma_entanglement_2020,YDing_PRA_2021,deng_sudden_2021,Sidek_Ent_2021}. A number of schemes that can either  advance, postpone or circumvent ESD have been demonstrated, such as the use of local unitary operations~\cite{Y_na__2007,Rau_EPL_2008,ali_manipulating_2008,MIH_2011,Singh_JOSAB_17,chathavalappil_schemes_2019}, quantum feedback~\cite{carvalho,Yamamoto_PRA_2008}, local Kerr medium~\cite{Qiang_2008}, off-resonant interactions~\cite{Chan_2009}, quantum interference~\cite{Das_PRA_2010}, weak measurement~\cite{Kim_Nat_Phys_2012}, non-Markovian environment~\cite{xu_esd,deng_sci_rep_2017,YMeng_PRA_2020,deng_sudden_2021},  decoherence free subspace~\cite{Bussandri_2020},  disorder in atom-cavity interactions~\cite{Ghosal_PRA_2020}, partial correlations between the consecutive actions of the correlated dephasing channel~\cite{Xue_Ent_2020}, etc. Further, 
some salient aspects of the dynamics of genuine multipartite entanglement have also been studied in this context in systems comprising three or more qubits~\cite{Rau_PRA_2014,kim_genuine_2016,Lu_2020,Kim_2021}.


However, a comprehensive study of the roles played by the specific interactions mentioned above and also the  initial state preparation, on the entanglement dynamics in hybrid quantum systems has not been reported in the literature. In this work, we use the double Jaynes-Cummings (DJC) model as the basic model, and examine how additional interactions  (such as the BS, DD and Ising  interactions) affect the subsequent dynamics. We specifically find that the Ising interaction leaves significant imprint on the entanglement dynamics. In certain parameter regimes this interaction leads to complete disappearance of ESD, while in  other regimes it can either significantly postpone the onset of ESD or reduce the time interval over which ESD is observed. We believe that these results can potentially solve some of the hurdles faced while executing several quantum tasks  where maintaining the entanglement for longer duration is essential. We have also  carried out a detailed investigation on the role of initial state preparation. For this purpose, we have examined various Bell-type entangled states for the two qubits, while for the pair of oscillators we have considered initial pure states (such as the ground  and excited Fock  states and the coherent state) as well as the thermal state.  In particular, we find  that while using coherent sources for the oscillators has a  positive effect for the entanglement in the CV subsystem,  this is at the expense of the extent of entanglement in the DV subsystem. 
For completeness,  we briefly discuss results on  the role of environmentally induced decoherence effects on the unitary entanglement dynamics. The effect of a zero temperature bath (vacuum as environment) as well as a thermal bath with finite temperature, on the unitary dynamics, is reported in this regard.

\section{\label{sec:model} The  heterogeneous quantum model}
The  Hamiltonian for the DJC model is given by
 \begin{align}
 \displaystyle H_{\text{DJC}} = H_{\text{free}} + H^{(1)}_{\text{JC}} + H^{(2)}_{\text{JC}} \,,
\label{eqn:ham_djcm}
\end{align}
where the free energy terms for the qubits and the oscillators are included in
\begin{align}
 \displaystyle H_{\text{free}} =  \tfrac{1}{2}\omega_{0} \left( \sigma_{z}^{(1)}  + \sigma_{z}^{(2)}\right) + \omega \left(a^{\dagger}a + b^{\dagger}b \right).
\label{eqn:ham_free}
\end{align}
Here, $\omega_{0}$  is the energy difference between the two states of the qubits and $\omega$ is the frequency of the oscillators. For simplicity, we set the frequencies of both  qubits (respectively, oscillators)  to be equal. Throughout the analysis we set the detuning $\Delta = \omega-\omega_0 $ to be zero (see, however, Section~\ref{sec:conclusion}, where we briefly comment on the role of imperfect frequency matching between the qubits and the oscillators on the dynamics, while a detailed analysis can be found in the Appendix).

Inclusion of  the beamsplitter, dipole-dipole and Ising interactions modifies the DJC Hamiltonian to 
\begin{align}
 H_{\text{tot}} = H_{\text{DJC}} + H_{\text{BS}} + H_{\text{DD}} + H_{\text{IS}}.
\label{eqn:ham_tot}
\end{align}
We examine how the various couplings affect the dynamics by  setting  $g_{\text{JC}}^{(1)}=g_{\text{JC}}^{(2)} = g_{\text{JC}}$ and work  with scaled dimensionless time $\tau = g_{\text{JC}}\, t$. The  effective coupling constants now become $r_{\text{b}} = g_{\text{b}}/g_{\text{JC}}$, $r_{\text{d}} = g_{\text{d}}/g_{\text{JC}}$ and $r_{\text{i}} = g_{\text{i}}/g_{\text{JC}}$.

We also study the roles played by two initial qubit-qubit entangled states given by  
\begin{align}
 \label{eqn:jcm_init1}
 \displaystyle \ket{\psi_{1}(0)} = \cos\phi \ket{e\,g} + \sin\phi \ket{g\,e},\\
 \displaystyle \ket{\psi_{2}(0)} = \cos\phi \ket{e\,e} + \sin\phi \ket{g\,g},
 \label{eqn:jcm_init2}
\end{align}
with $\phi=\tfrac{\pi}{4}$ corresponding to the maximally entangled Bell states. The initial oscillator states considered  are the ground state $\ket{0}$,  the excited Fock states $\ket{n}$, the standard coherent state (CS) $\ket{\alpha}$ with intensity  $\vert\alpha\vert^2$, and the thermal state $\rho_{\text{th}}$ with mean thermal energy $\bar{n}$. In the Fock basis $\{\ket{n}\}$, the CS and the thermal state are, respectively,
\begin{align}
 \displaystyle \ket{\alpha} &= e^{-\frac{1}{2}\vert\alpha\vert^2}\sum_{n=0}^{\infty} \frac{\alpha^{n}}{\sqrt{n!}}  \ket{n},\\
 \displaystyle \rho_{\text{th}} &= \frac{1}{1+\bar{n}}\sum_{n=0}^{\infty} \left(\frac{\bar{n}}{1+\bar{n}}\right)^{n}  \ket{n}\bra{n}.
\end{align}
 
 The extent of qubit-qubit entanglement is measured in terms of the concurrence $C$~\cite{wooters}. For a pair of qubits with density matrix $\rho$, 
\begin{equation}
 \displaystyle C = \max\{0, \lambda_1 -  \lambda_2 -  \lambda_3 - \lambda_4\},
\label{eqn:conc_def}
\end{equation}
 where $\lambda_{j}$'s ($j=1,\cdots,4$) are obtained by taking the square roots of the eigenvalues of the non-Hermitian matrix $\mathcal{R} = \rho \tilde{\rho}$. Here,   
 $ \tilde{\rho} = (\sigma_{y}\otimes\sigma_{y}) \rho^{*} (\sigma_{y}\otimes\sigma_{y})$,  and $\rho^{*}$ is the complex conjugate of $\rho$ and $\sigma_{y}$ is  the usual Pauli matrix.
 The extent of oscillator-oscillator entanglement, on the other hand, is measured in terms of the  logarithmic negativity (LN). This is equal to $\log_{2}\vert\vert\rho^{\Gamma A}\vert\vert_{1}$, where $\Gamma A$ is the partial transpose operation and $\vert\vert\rho\vert\vert_{1}$ denotes the trace norm of $\rho$~\cite{plenio,vidal}.  


\section{\label{sec:result} Results}
 The results presented in this article do not necessarily rely on a specific physical  implementation of the Hamiltonian $H_{\text{tot}}$. However, for details regarding possible experimental implementations of the various parts of $H_{\text{tot}}$ (Eq.~\eqref{eqn:ham_tot}), we refer the reader to the Appendix.

\subsection{Oscillators prepared in the ground state}
\label{sec:gs}
 For both oscillators prepared initially in their respective ground states, the problem simplifies considerably. For ready reference, we first recall the key inferences resulting from the DJC model~\cite{Y_na__2006} (see Appendix). Firstly, for initial entangled states $\ket{\psi_1(0)}$  and $\ket{\psi_2(0)}$  (Eq.~\eqref{eqn:jcm_init1} and Eq.~\eqref{eqn:jcm_init2}), the state vectors of the full system at a later time $\tau$, can be written as
\begin{align}
 \label{eqn:psi1t_gs}
 \ket{\Psi_{1}(\tau)} &= x_{1}(\tau) \ket{e\, g\, 0\, 0} + x_{2}(\tau) \ket{g\, e\, 0\, 0} + x_{3}(\tau) \ket{g\, g\, 1\, 0} \nonumber \\[2pt]
                      &\quad + x_{4}(\tau) \ket{g\, g\, 0\, 1},\\ 
 \vspace{2ex}
 \ket{\Psi_{2}(\tau)} &= y_{1}(\tau)\ket{e\, e\, 0\, 0} + y_{2}(\tau)\ket{g\, g\, 1\, 1} + y_{3}(\tau)\ket{e\, g\, 0\, 1}\nonumber \\[2pt]
                     &\quad +y_{4}(\tau)\ket{g\, e\, 1\, 0} + y_{5}(\tau)\ket{g\, g\, 0\, 0}.
\label{eqn:psi2t_gs} 
\end{align}
The oscillators can be treated as qubits, as only two energy levels ($\ket{0}$ and  $\ket{1}$) for each, are relevant for the dynamics, and the system effectively reduces to a 4-qubit hybrid system. In addition to qubit-qubit entanglement, the extent of oscillator-oscillator entanglement can also be measured in terms of concurrence, which for the initial entangled states $\ket{\psi_{1}(0)}$ are (see Appendix)
\begin{align}
\label{concq1_djc}
  {C_{\text{q}}}_1 &= \vert\sin(2\phi)\vert \cos^2\tau,\\
\label{conco1_djc}
  {C_{\text{o}}}_1 &= \vert\sin(2\phi)\vert \sin^2\tau,
\end{align}
while  for  $\ket{\psi_{2}(0)}$ these are given by
\begin{align}
\label{concq2_djc}
    {C_{\text{q}}}_2 &= \text{max}\left\{0,\, \cos^2\tau \left(\vert\sin(2\phi)\vert - 2\cos^2\phi \sin^2\tau\right)\right\}, \\
\label{conco2_djc}
   {C_{\text{o}}}_2 &= \text{max}\left\{0,\,  \sin^2\tau \left(\vert\sin(2\phi)\vert - 2\cos^2\phi \cos^2\tau\right)\right\}.
\end{align}
Firstly, we see that for initial state $\ket{\psi_1(0)}$, the qubit-qubit (similarly, the oscillator-oscillator) subsystem exhibits  perfect oscillatory entanglement dynamics with maxima (minima) at $\tau = l\, \pi$ and minima (maxima) at $\tau = \left(l+\tfrac{1}{2}\right)\, \pi,\,\,l = 0, 1, 2, \cdots$.  
The maximum value of the entanglement is set by the degree of entanglement of the initial state (i.e., $\phi$). We emphasise that the  subsequent entanglement is generated because of the initial qubit-qubit  entanglement  and not because of direct interaction between them. More importantly, ESD  is absent. In  contrast, the dynamics gets significantly modified, both quantitatively and qualitatively, for initial entangled state $\ket{\psi_{2}(0)}$. Depending on the choice of $\phi$, we see sudden death (whenever $2\cos^2\phi\sin^2\tau > \vert\sin(2\phi)\vert$) and subsequent birth of qubit-qubit entanglement (similarly for the oscillator-oscillator subsystem), again in an oscillatory manner.

\begin{figure}[ht!]
  \centering
  \includegraphics[height=12.0cm,width=0.48\textwidth]{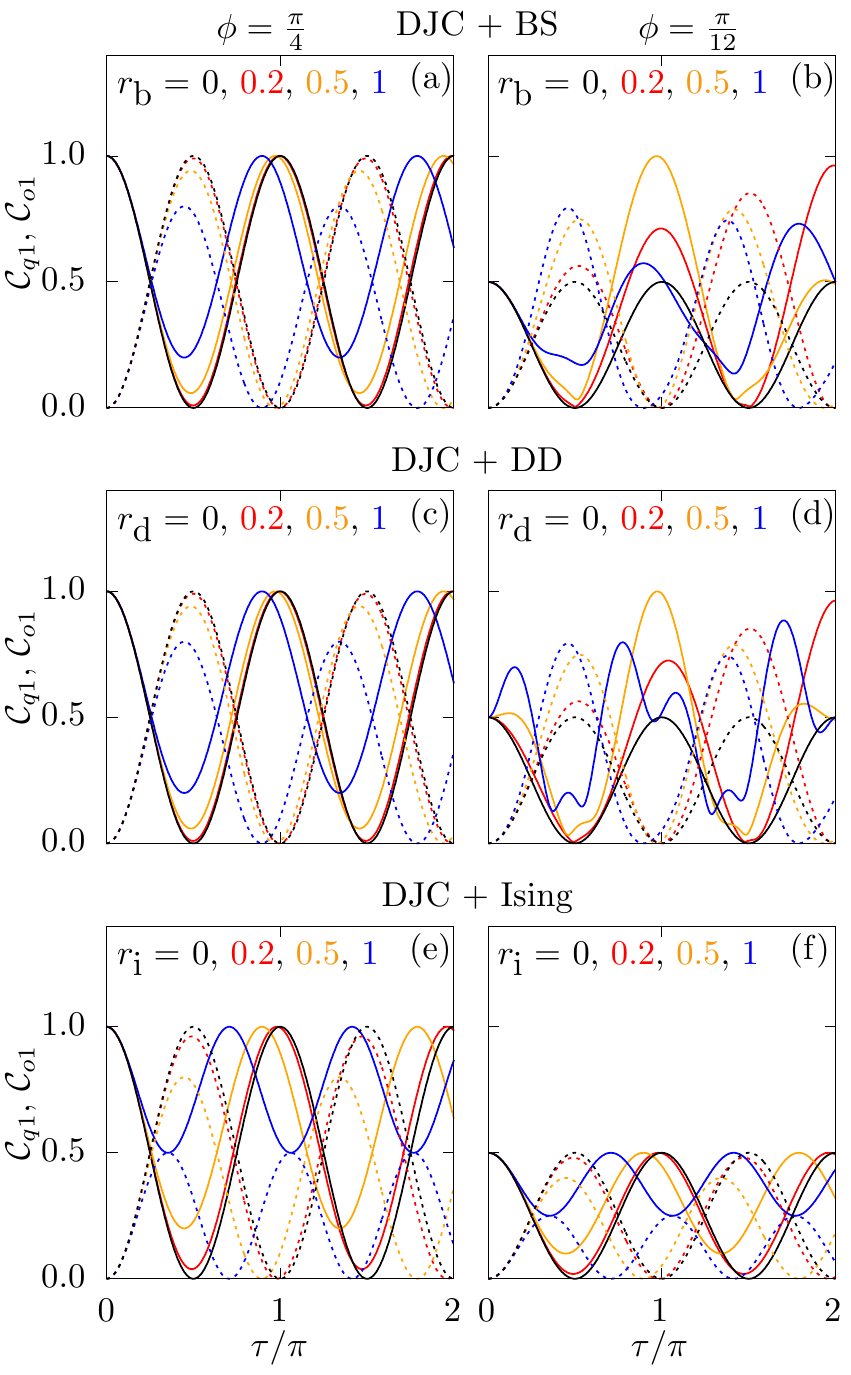}
  \vspace{-2.50ex} 
  \caption{Concurrence for the qubit-qubit (solid) and oscillator-oscillator (dashed) subsystems as a function of scaled time $\tau$ for the DJC model  with additional beamsplitter (top panel), dipole-dipole (center panel) and Ising (bottom panel) interactions having different coupling strengths (indicated in each plots).  Initially, qubits are prepared in entangled state $\ket{\psi_{1}(0)}$ with $\phi = \tfrac{\pi}{4}$ (left panel) and $\tfrac{\pi}{12}$ (right panel), while  both oscillators are prepared in  the ground states. With additional BS and dipole-dipole interactions, perfect oscillatory dynamics is lost due to the continuous interference effects for $\phi = \tfrac{\pi}{12}$. Also, maximum qubit-qubit entanglement is reached in both cases although the initial entanglement is significantly less.}
  \label{fig:conc_jcm_all_psi1}
\end{figure}

\begin{figure}[ht!]
  \centering
  \includegraphics[height=12.0cm,width=0.48\textwidth]{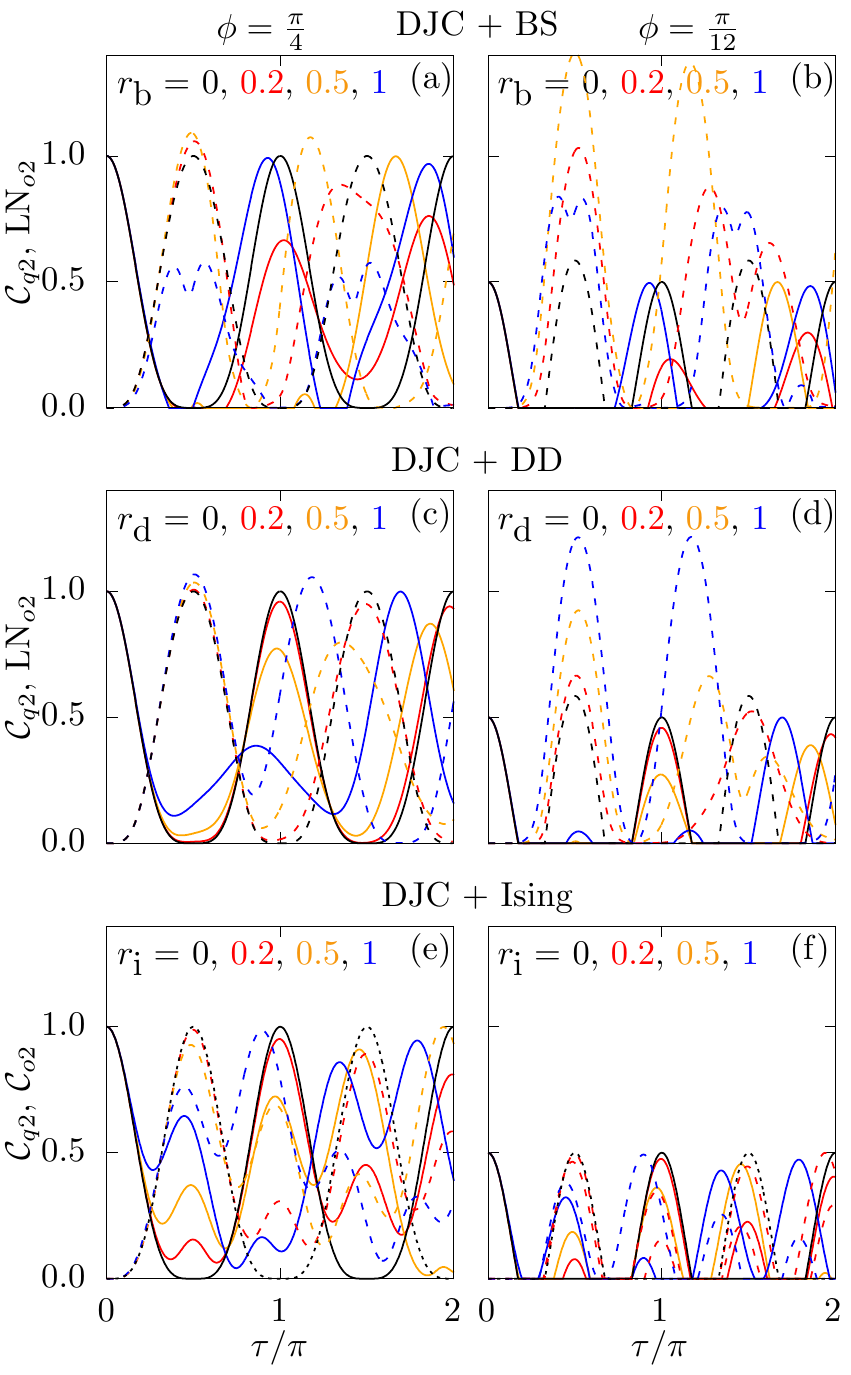}\\
  \vspace{-1.0ex}
  \caption{Entanglement dynamics for the double Jaynes-Cummings model as a function of scaled time $\tau$ with additional beamsplitter (top panel), dipole-dipole (center panel) and Ising (bottom panel) interactions having different coupling strengths (indicated in each plots). Initially, qubits are in entangled state $\ket{\psi_{2}(0)}$ with $\phi = \tfrac{\pi}{4}$ (left panel) and $\tfrac{\pi}{12}$ (right panel), while  both oscillators are in  the ground states. The solid (dashed) lines correspond to the qubit-qubit (oscillator-oscillator) concurrence (LN (panels a-d) and concurrence (panels e, f)).}
  \label{fig:conc_jcm_all_psi2}
\end{figure}

Bearing these results in mind, we now investigate  the crucial roles played by additional BS, DD and Ising interactions on the entanglement dynamics (Figs.~\ref{fig:conc_jcm_all_psi1} and \ref{fig:conc_jcm_all_psi2}). For an initial entangled  state $\ket{\psi_{1}(0)}$, the oscillators  continue to populate only the first two energy levels during temporal evolution, and hence they can still be treated as qubits. The state vector has the same form as in Eq.~\eqref{eqn:psi1t_gs} but with modified time-dependent coefficients (Appendix). The qubit-qubit concurrence for this case is
\begin{align}
 {C_{\text{q}}}_1 = &2 \Big\{\left(\cos^2\phi\,\,  g^2(\tau_1,\tau_2) + \sin^2\phi\,\, h^2(\tau_1,\tau_2)\right) \nonumber\\
      &\,\times \left(\sin^2\phi\,\, g^2(\tau_1,\tau_2) + \cos^2\phi\,\, h^2(\tau_1,\tau_2)\right) \Big\}^{\frac{1}{2}},
 \label{eqn:qbt_conc1_bs_gs}
\end{align}
where
\begin{subequations}
 \label{eqn:ght_bs}
\begin{align}
 \label{eqn:gt}
    g(\tau_1,\tau_2) = \cos\tau_1\,\cos\tau_2 + \frac{\tau_1}{\tau_2}\, \sin\tau_1 \sin\tau_2, \\
 \label{eqn:ht}
    h(\tau_1,\tau_2) = \sin\tau_1\,\cos\tau_2 - \frac{\tau_1}{\tau_2}\, \cos\tau_1\sin\tau_2,
\end{align}
\end{subequations}
with $\tau_1 = \tfrac{r_{\text{b}}}{2}\, \tau$, $\tau_2 = \sqrt{1+\tfrac{r_{\text{b}}^2}{4}}\, \tau$. Note that $\tau^2  = \tau_2^2 - \tau_1^2$. The BS interaction introduces an additional time scale in the system. From Eq.~\eqref{eqn:qbt_conc1_bs_gs} it is clear that due to  continuous interference between oscillations of two different frequencies $\tau_1$ and $\tau_2$, perfect oscillations are lost (solid curves in Fig.~\ref{fig:conc_jcm_all_psi1}(b)), leading to complex dynamical evolution depending on the interaction strength $r_\text{b}$.  However, for $\phi=\frac{\pi}{4}$, Eq.~\eqref{eqn:qbt_conc1_bs_gs} simplifies remarkably, and we get  ${C_{\text{q}}}_1 = 1- \frac{\tau^2}{\tau_2^2}\sin^2\tau_2$; exact oscillations arise (Fig.~\ref{fig:conc_jcm_all_psi1}(a)).  The period of oscillations, however, is modified by the strength of the coupling,~$r_{\text{b}}$. From Eq.~\eqref{eqn:qbt_conc1_bs_gs}, it is also clear why with increasing $r_{\text{b}}$ the minimum value of concurrence moves upwards from zero. Another interesting observation pertains to the maximum value of the entanglement $C_1^{\text{max}}$. In the absence of the BS interaction,  ${C_{\text{q}}}_1^{\text{max}}=1$ only when $\phi = \frac{\pi}{4}$ (see Eq.~\eqref{concq1_djc}). With the BS coupling on, this constraint on  ${C_{\text{q}}}_1^{\text{max}}$ is absent, and ${C_{\text{q}}}_1^{\text{max}}=1$ is attained for other values of $\phi$ as well (yellow curve in Fig.~\ref{fig:conc_jcm_all_psi1}(b)). Thus, even if the initial two-qubit state is weakly entangled, we can always produce a maximally entangled two-qubit state using the additional BS interaction. 

On the other hand, the oscillator-oscillator concurrence for a generic value of $\phi$ has a relatively simple form, compared to Eq.~\eqref{eqn:qbt_conc1_bs_gs}, and is given by 
\begin{align}
    {C_{\text{o}}}_1 = \frac{2\tau^2}{\tau_2^2}\,& \sin^2\tau_2 \Big\{\left(\cos^2\phi \cos^2\tau_1 + \sin^2\phi \sin^2\tau_1\right) \nonumber \\
    &\quad\left(\cos^2\phi \sin^2\tau_1 + \sin^2\phi \cos^2\tau_1\right)\Big\}^{\frac{1}{2}}.
 \label{eqn:osc_conc1_bs}
\end{align}
This reduces to ${C_{\text{o}}}_1 = \frac{\tau^2}{\tau_2^2}\, \sin^2\tau_2 $ for $\phi = \frac{\pi}{4}$ (Appendix), and hence  perfect oscillations arise. For other values of $\phi$, the oscillations are modulated by the multiplying factor. Further, the oscillations always start at zero, in contrast to  ${C_{\text{q}}}_1$.

With addition of the dipole-dipole interaction to the DJC Hamiltonian, the equation for the evolution of the state vector 
 is  given by  Eq.~\eqref{eqn:psi1t_gs}. Further, ${C_{\text{q}}}_1$ has the same analytical form as that in Eq.~\eqref{eqn:qbt_conc1_bs_gs}  (with the identification $r_{\text{b}} = r_{\text{d}}$), and with 
\begin{subequations}
\label{eqn:ght_dd}
\begin{align}
    g(\tau_1,\tau_2) = \cos\tau_1\,\cos\tau_2 - \frac{\tau_1}{\tau_2}\, \sin\tau_1 \sin\tau_2, \\
    h(\tau_1,\tau_2) = \sin\tau_1\,\cos\tau_2 + \frac{\tau_1}{\tau_2}\, \cos\tau_1\sin\tau_2.
\end{align}
\end{subequations}
Note the subtle changes in sign in Eq.~\eqref{eqn:ght_bs} and Eq.~\eqref{eqn:ght_dd}.  Since the contribution from the oscillators is effectively like that from  qubits, the BS interaction resembles the DD interaction. This is affirmed by  the analytical expressions and augmented by the numerical computations (see Figs.~\ref{fig:conc_jcm_all_psi1}(a--d)).



In the presence of additional Ising interaction  to the DJC Hamiltonian, the qubit-qubit and the oscillator-oscillator concurrences are found to be 
\begin{align}
 {C_{\text{q}}}_1 &= \vert\sin(2\phi)\vert\left(1- \frac{\tau^2}{\tau_1^2}\,\sin^2\tau_1\right)\,, \\
 {C_{\text{o}}}_1 &=  \vert\sin(2\phi)\vert\, \frac{\tau^2}{\tau_1^2} \sin^2\tau_1\,,
\end{align}
where $\tau_1 = \sqrt{1+r_{\text{i}}^2}\, \tau$ (Appendix).
Therefore, for all values of $\phi$, ${C_{\text{q}}}_1$ (similarly, ${C_{\text{o}}}_1$) continues to exhibit perfect oscillations ((see Figs.~\ref{fig:conc_jcm_all_psi1}(e,f))) which are bounded by $\frac{r_{\text{i}}^2}{1+r_{\text{i}}^2}\vert\sin(2\phi)\vert$ and $\vert\sin(2\phi)\vert$ (similarly, 0 and $\vert\sin(2\phi)\vert$). The period of oscillations gets modulated by the Ising interaction strength, $r_\textrm{i}$.  Comparing the expressions for concurrence when $\phi=\frac{\pi}{4}$, we see that although the qualitative behaviour is identical, the effect of the additional Ising interaction is much stronger compared to the BS/DD interactions. In both cases the frequencies are modulated by a factor of $\sqrt{1+\frac{r_{\text{b}}^2}{4}}$ and $\sqrt{1+r_{\text{i}}^2}$, respectively.

Finally, we note that only for an additional Ising interaction  the `specific pairwise entanglement' ($C_{\text{qo}} = {C_{\text{q}}}_1+{C_{\text{o}}}_1$) is conserved (i.e., $C_{\text{qo}} = \vert\sin(2\phi)\vert$, similar to the DJC dynamics) while for the BS or DD case, this holds only  if $\phi=\frac{\pi}{4}$.

We now consider the two-qubit entangled initial state $\ket{\psi_{2}(0)}$ that leads to ESD.  We note that after including an additional BS or DD interaction to the DJC Hamiltonian, the state vector of the full system cannot  be written as in Eq.~\eqref{eqn:psi2t_gs}. Instead it has  the following form
\begin{align}
 \ket{\Psi_{2}(\tau)} &= y_{1}\ket{e\, e\, 0\, 0} + y_{2}\ket{g\, g\, 1\, 1} + y_{3}\ket{e\, g\, 0\, 1}\nonumber \\
           &\,\, +y_{4}\ket{g\, e\, 1\, 0} + y_{5}\ket{g\, g\, 0\, 0} + y_{6}\ket{g\, g\, 2\, 0}\nonumber \\
           &\,\,+ y_{7}\ket{g\, g\, 0\, 2} + y_{8}\ket{e\, g\, 1\, 0} + y_{9}\ket{g\, e\, 0\, 1}.
\label{eqn:psi2t_bs} 
\end{align}
 In contrast to the earlier cases we have investigated here, the oscillator's contributions arise  also from $\ket{2}$, apart from  $\ket{0}$ and $\ket{1}$. Therefore, the oscillators can no longer be treated as qubits and we will use LN as the measure of oscillator-oscillator entanglement. 
For additional BS as well as additional DD interactions, exact analytical expressions for these nine time dependent coefficients cases are obtained (Appendix). It is interesting to note that the qubit-qubit density matrix has the appearance of an $X$-state (Appendix).  
In both these cases, we have extracted the two time scales that control the dynamics (Appendix). For an additional BS coupling, these are $\tau_{\pm} = \sqrt{\frac{1}{2}(5r_b^2+6\pm\Gamma_b)}\, t$, where $\Gamma_b = \sqrt{9r_b^4 + 60r_b^2+4}$.  For an additional DD coupling, $\tau_\pm = \sqrt{\frac{1}{2}(r_d^2+6\pm\Gamma_d)}\, t$, where $\Gamma_d=\sqrt{r_d^4+12 r_d^2+4}$. The complex algebraic forms result in significantly complex entanglement dynamics (for all values of $\phi$), as can be seen in Figs.~\ref{fig:conc_jcm_all_psi2}(a--d). When $r_{\text{b}}=1$, we observe a significant reduction in the  duration of ESD, for both values of $\phi$ (Figs.~\ref{fig:conc_jcm_all_psi2}(a,\,b)). We also note the oscillator-oscillator entanglement going past the value 1 in both cases, signalling the contributions coming from $\ket{2}$. The entanglement reaches its maximum value for $r_\text{b}=0.5$ and $\phi=\frac{\pi}{12}$ (yellow dashed curve in Fig.~\ref{fig:conc_jcm_all_psi2}(b)). 
The qualitative features remain similar for the additional dipole-dipole coupling, except that the oscillator-oscillator entanglement reaches its maximum value when $r_\text{d}=1$ and $\phi=\frac{\pi}{12}$ (blue dashed curve in Fig.~\ref{fig:conc_jcm_all_psi2}(d)).

In the presence of an additional Ising interaction to the DJC Hamiltonian, the equation for the state vector remains simple, as in Eq.~\eqref{eqn:psi2t_gs}, and only the coefficients get modified (Appendix).
The qubit-qubit and oscillator-oscillator concurrence in this case, can be expressed as ${C_{\text{q}}}_2 = \text{max}\left\{0,\,f_{+}(\tau)\right\}$ and ${C_{\text{o}}}_2 = \text{max}\left\{0,\,f_{-}(\tau)\right\}$, where
\begin{align}
  f_{\pm}(\tau) &= -\frac{2\tau^2}{\tau_2^2} \cos^2\phi\sin^2\tau_2 + \frac{1}{2}\vert\sin(2\phi)\vert\nonumber \\
        &\qquad\times \sqrt{\left(\cos\tau_1\pm\cos\tau_2\right)^2+\left(\sin\tau_1\pm\frac{\tau_1}{\tau_2}\sin \tau_2\right)^2},
 \end{align}
with $\tau_1 = r_{\text{i}} \tau$ and $\tau_2 = \sqrt{4+r_{\text{i}}^2} \, \tau$. 

Unlike for the initial state $\ket{\psi_1(0)}$, here there is continuous interference between oscillations  at two different frequencies. Hence, apart from the frequency of oscillations getting modulated, the interference also results in disappearance of ESD for $\phi=\frac{\pi}{4}$ (Fig.~\ref{fig:conc_jcm_all_psi2}(e)), while for other values of $\phi$ the duration of ESD becomes shorter  as the Ising interaction strength is increased (Fig.~\ref{fig:conc_jcm_all_psi2}(f)).


Finally, we note that the `specific pairwise entanglement' $C_{\text{qo}}$ in all these three cases is not conserved, in contrst to  the case of the DJC model in the absence of any additional interactions. In fact, it has been shown~\cite{Lopez_PRL_2008,Rau_PRA_2014} that the exact times of ESD ($t_{\text{ESD}}$) and entanglement sudden birth ESB ($t_{\text{ESB}}$) can be manipulated in such a way that $t_{\text{ESD}}$ can be greater than, equal to or less than  $t_{\text{ESB}}$. In particular, there can be a time window where neither the qubits nor the cavities are entangled among themselves. This time window is shown to have a profound effect on the dynamics of genuine multipartite entanglement~\cite{Rau_PRA_2014}.


\subsection{Oscillators prepared in excited Fock states}
Now, we assume that the oscillators are initially in arbitrary Fock states $\ket{n}$ and $\ket{m}$, respectively. 
We consider the DJC model excluding all the additional interactions. Corresponding to  initial entangled states $\ket{\psi_1(0)}$ and $\ket{\psi_2(0)}$, the dynamics involves  eight time-dependent coefficients (see Appendix, where exact analytical expressions are derived for the state vectors). The qubit-qubit entanglement can be written as ${C_{\text{q}}}_i =  \text{max}\left\{0,\, f_i(\tau)\right\}$ ($i=1,\,2$ corresponding to the two initial states ), where (see Appendix)
\begin{align}
\label{eqn:conc1_fock} 
 &f_1(\tau) = \vert\sin(2\phi)\vert \cos\tau_1  \cos \tau_2\Big\{\cos\tau_3\cos\tau_4 \nonumber  - \sin\tau_3  \sin\tau_4\\
    &\,\,\times\sqrt{\cos^2\phi \sin^2\tau_1\cos^2\tau_4  + \sin^2\phi \sin^2\tau_2 \cos^2\tau_3} \Big\}\,,
\end{align}
and
\begin{align}
\label{eqn:conc2_fock} 
  f_2&(\tau) =\vert\sin(2\phi)\vert \cos\tau_1  \cos \tau_2 \cos\tau_3\cos\tau_4 \nonumber  \\
    &-2 \Big\{\left(\cos^2\phi \sin^2\tau_1\cos^2\tau_2  + \sin^2\phi  \cos^2\tau_3\sin^2\tau_4\right)^2 \nonumber\\
    &+ \left(\cos^2\phi\cos^2\tau_1 \sin^2\tau_2  + \sin^2\phi\sin ^2\tau_3 \cos^2\tau_4 \right)^2 \Big\}^{\frac{1}{2}},
\end{align}
respectively. Here $\tau_1 = \sqrt{n+1}\, \tau$, $\tau_2 = \sqrt{m+1}\, \tau$, $\tau_3 = \sqrt{n}\, \tau$ and $\tau_4 = \sqrt{m}\, \tau$. Eq.~\eqref{eqn:conc1_fock}  clearly shows that ESD is absent iff $n=m=0$. For all other finite values of $n$ and $m$, ESD is present even for the initial state $\ket{\psi_1(0)}$.
Further, from Eq.~\eqref{eqn:conc1_fock} and Eq.~\eqref{eqn:conc2_fock} it is clear that the dynamics is  fairly complicated even in the absence of any additional interactions as there are four time scales in the problem.

\begin{figure}[ht!]
  \centering
    \includegraphics[height=7.0cm,width=0.48\textwidth]{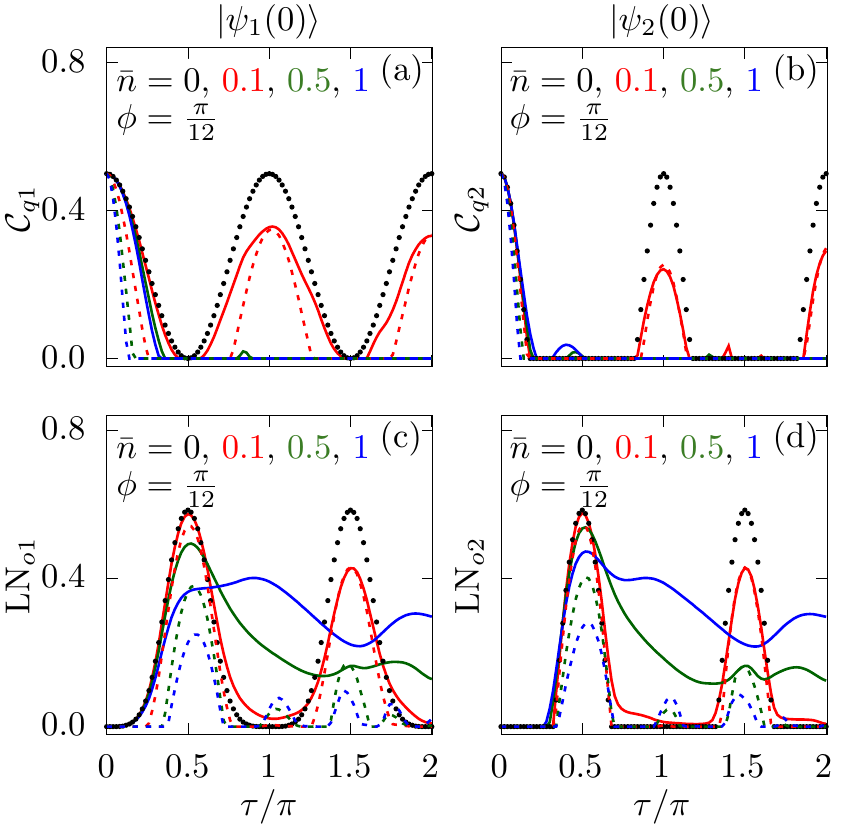}
    \vspace{-2.0ex}
  \caption{Entanglement dynamics of the double Jaynes-Cummings model with initial coherent oscillator states $\ket{\alpha}$ (solid) and thermal oscillator states $\rho_{\text{th}}$ (dashed) with strength $\bar{n}=\vert\alpha\vert^2$: qubit-qubit concurrence (top panel) and LN for the oscillator-oscillator subsystem (bottom panel) are plotted as a function of  scaled time $\tau$. The initial qubit-qubit entangled states are  $\ket{\psi_1(0)}$ (left panel) and $\ket{\psi_2(0)}$ (right panel), with $\phi = \tfrac{\pi}{12}$. The black dotted curves in each plot indicate the dynamics corresponding to initial ground state oscillators.}
  \label{fig:conc_cs_time}
\end{figure}

\subsection{Oscillators prepared in coherent and thermal states}
 We now  investigate the effects of initial coherent or thermal oscillator states on the entanglement dynamics. The exact analytical expressions for the state vectors for initial  $\ket{\psi_1(0)}$ and $\ket{\psi_2(0)}$ and  CS $\ket{\alpha}$ can be found in the Appendix. It can be seen that the reduced density matrices involve multiple summations over the eight time-dependent coefficients with multiplicative factors which cannot be expressed in simple algebraic forms. Hence, we resort to numerical analysis for inferences. Based on   Fig.~\ref{fig:conc_cs_time}, two interesting observations can be made on the concurrence. Firstly, the perfect oscillatory dynamics disappears rapidly with increase in $\vert\alpha\vert^2$ for both qubit-qubit initial entangled  states $\ket{\psi_1(0)}$ and $\ket{\psi_2(0)}$. We have  verified that this behaviour is independent of the choice of $\phi$, as long as the qubits are entangled. Secondly, unlike the earlier instance when both oscillators were initially in their respective ground states, ESD now appears even for the  initial state $\ket{\psi_1(0)}$.  Both these observations can be substantiated using our earlier results on initial Fock states for the oscillators. We have already noticed the existence of four time scales corresponding to  initial Fock states.  As we increase $\vert\alpha\vert^2$ more excited  energy levels in the oscillator ladder contribute. This leads to stronger interference effects that eventually destroys the perfect oscillatory dynamics. Also, the latter observation is expected as we have already seen ESD appearing even for $\ket{\psi_1(0)}$ with the  initial Fock state oscillators.
Further, the time duration over which ESD occurs increases significantly with increase in $\vert\alpha\vert^2$. The most striking difference however  is observed in the oscillator-oscillator entanglement dynamics. In this case, ESD disappears completely with increase in $\vert\alpha\vert^2$ for $\phi = \tfrac{\pi}{4}$.  For  other values of $\phi$,  ESD is absent only after some time elapses.  

%
\begin{figure*}[ht!]
  \centering
    \includegraphics[height=10.5cm,width=\textwidth]{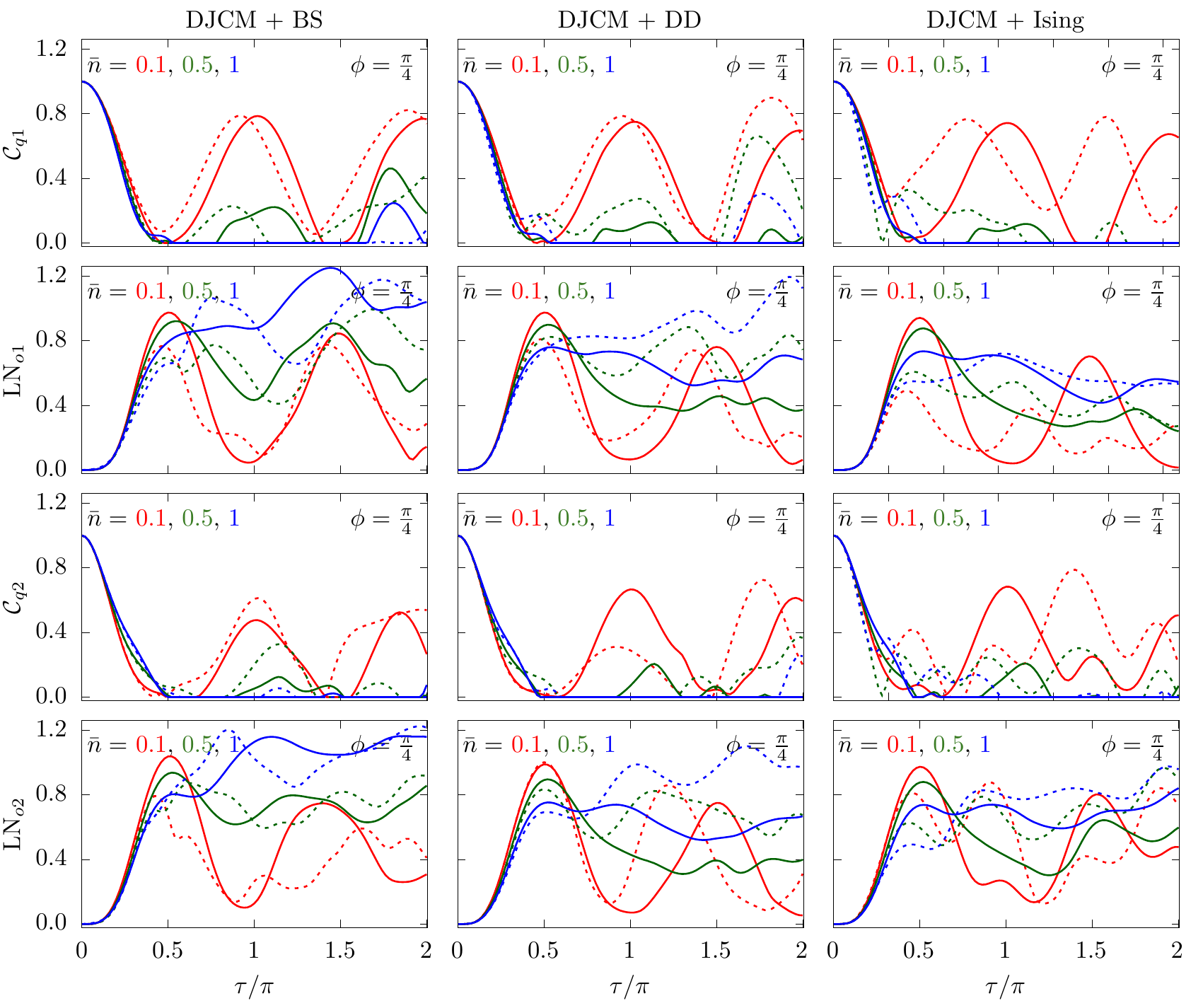}\\
    \vspace{-1.5ex}
  \caption{The effect of additional interactions on the entanglement dynamics is shown as a function of scaled time $\tau$. First and  third rows (respectively, second and fourth rows) are concurrence (LN) for the qubit-qubit (respectively, oscillator-oscillator). Initial qubit-qubit entangled states are  $\ket{\psi_1(0)}$ (first and second rows) and  $\ket{\psi_2(0)}$ (third and fourth rows), $\phi = \tfrac{\pi}{4}$. Initial oscillators are CS $\ket{\alpha}$ with $\vert\alpha\vert^2=0.1$ (red), 0.5 (green) and 1 (blue). The solid (dashed) lines correspond to scaled interaction strengths ($r_{\text{b}}$, $r_{\text{d}}$ and $r_{\text{i}}$) equal to $0.2$ ($0.8$).}
  \label{fig:conc_cs_t_all}
\end{figure*}

Next, we examine the changes that arise due to relaxation of initial purity of the oscillator states. For initial thermal resources for the oscillators the entanglement dynamics is shown  by the dashed curves in (Fig.~\ref{fig:conc_cs_time}). Without loss of generality we have assumed that both the oscillators have equal thermal energy, characterised by $\bar{n}$. We can identify two striking differences in the dynamics between the coherent and thermal oscillator states.
Firstly, for the qubit-qubit subsystem, we see that the onset of ESD is much quicker in the case  of initial thermal oscillators. Secondly, and the more important difference is that the duration of ESD does not lessen  as dramatically and rapidly for the oscillator-pair subsystem, compared to the case of initial coherent resources discussed earlier. 

 We observe in passing that although perfect oscillations are lost at short times and decay of entanglement can be seen for both these initial states, the dynamics, in fact, continues to be oscillatory  at long times albeit with complex, imperfect oscillations (Appendix).

\begin{figure*}[ht!]
  \centering
    \includegraphics[height=10.5cm,width=\textwidth]{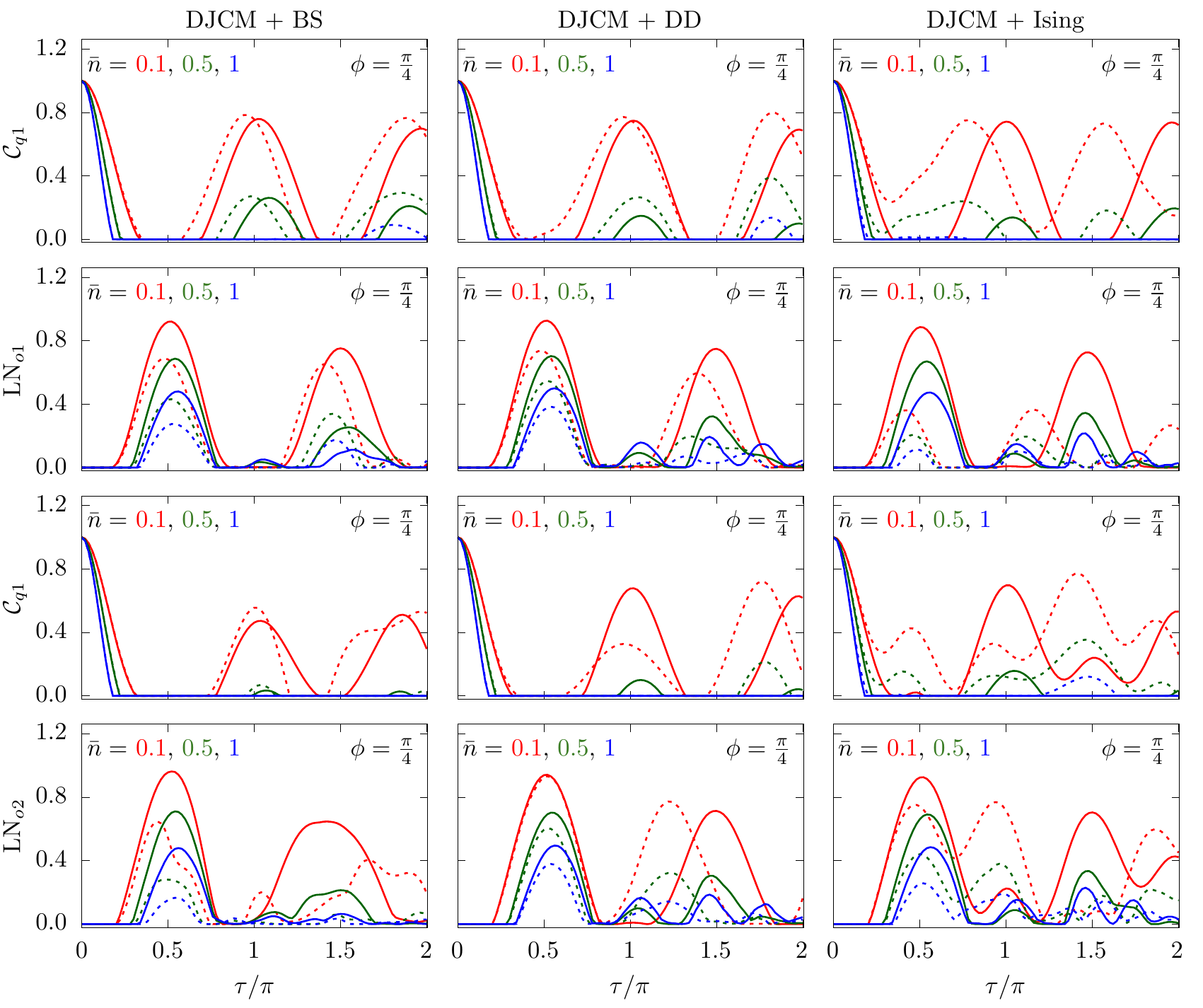}\\
    \vspace{-1.5ex}
  \caption{Similar to Fig.~\ref{fig:conc_cs_t_all} but now both the oscillators are initially  in thermal states  with equal thermal energy $\bar{n}=0.1$ (red), 0.5 (green) and 1.0 (blue). }
  \label{fig:conc_th_t_all}
\end{figure*}

\subsection{Effect of additional interactions for initial coherent or thermal oscillators}
In Figs.~\ref{fig:conc_cs_t_all} and \ref{fig:conc_th_t_all}, we see some of the effects of additional interactions on the entanglement dynamics for different values of $\vert\alpha\vert^2$ and $\bar{n}$, corresponding to initial CS and thermal states of  the oscillators. In both cases, we have verified  that (as expected), for extremely small values of $\vert\alpha\vert^2$ and $\bar{n}$, the changes in the dynamics are quite similar to that of initial ground state oscillators. The impact of these additional interactions  becomes significant for relatively higher values of $\vert\alpha\vert^2$ and $\bar{n}$. 
For instance, consider the qubit-qubit entanglement dynamics for $\ket{\psi_1(0)}$. We have verified that if we include only the linear beamsplitter coupling, then ESD for  $0<\vert\alpha\vert^2 \lessapprox 0.3$ disappears completely as  $r_{\text{b}}$ is increased gradually from 0 to 1, for all values of $\phi$. As  $\vert\alpha\vert^2$ is increased beyond 0.3, ESD reappears.  The time duration of ESD however reduces as $r_{\text{b}}$ is increased gradually from 0 to 1, irrespective of the value of $\phi$. 
In contrast, when the qubits are initially in the state $\ket{\psi_2(0)}$,  for  $0<\vert\alpha\vert^2 \leq 0.3$,  the time duration of ESD reduces as the value of $r_{\text{b}}$ is increased. We observe that this time interval is significantly less for $\phi = \tfrac{\pi}{4}$, compared to other values of $\phi$.
Now, for the oscillator-oscillator entanglement dynamics, although there are quantitative differences that appear with increasing $r_{\text{b}}$ for both  initial states $\ket{\psi_1(0)}$ and $\ket{\psi_2(0)}$, we have verified that the overall qualitative behaviour still remains similar to that when $r_{\text{b}}=0$. 

The inclusion of an additional dipole-dipole interaction has qualitatively similar impact  as that of the BS term, on the entanglement dynamics, except for some subtle modifications. For instance, complete disappearance in the qubit-qubit ESD now happens in the range $0<\vert\alpha\vert^2 \lessapprox 0.25$ for initial $\ket{\psi_1(0)}$. Further, for higher $\vert\alpha\vert^2$ the onset of ESD is relatively delayed.

The changes observed when an additional Ising coupling is added over and above the  basic double JC interaction are the following.
First, consider the qubit-qubit entanglement dynamics. We notice a significant delay in the occurrence of ESD for both initial $\ket{\psi_1(0)}$ and $\ket{\psi_2(0)}$ as the value of $r_{\text{i}}$ is increased from 0 to 1. The effect is more evident  for $\ket{\psi_2(0)}$, for relatively smaller values of $\vert\alpha\vert^2$. In particular, for $\phi=\tfrac{\pi}{4}$, ESD is completely absent for a substantial interval of time, if $\vert\alpha\vert^2\lessapprox 0.8$, in contrast to other values of $\phi$ and for initial state $\ket{\psi_1(0)}$. This is in contrast to the earlier result that for the initial ground state oscillators we observe ESD  only for initial $\ket{\psi_2(0)}$. 
Once again, there is no ESD in the oscillator-oscillator subsystem and the overall extent of entanglement for all the three cases increases with increase in $\vert\alpha\vert^2$ and coupling strength. Apart from this  feature, the qualitative behaviour is similar to the case where no additional interaction is included.

From Fig.~\ref{fig:conc_th_t_all} we can understand  the roles played by different additional interactions when  initial thermal state of the oscillators are considered, and compare the results with initial coherent state oscillators. In the presence of an additional BS coupling, the time interval of ESD  for the qubit-qubit subsystem for initial $\ket{\psi_1(0)}$ and $\ket{\psi_2(0)}$, reduces when $\bar{n}$ is smaller, while for larger $\bar{n}$ increasing $r_{\text{b}}$ does not necessarily affect the time duration of ESD substantially. This reduction in the duration of ESD is much more prominent for an additional DD coupling when compared with the additional BS coupling. Once again, we find that the additional Ising interaction has stronger impact on the entanglement dynamics, when compared with an additional  BS or DD interaction. 
Further, one important difference that is clearly seen by comparing Figs.~\ref{fig:conc_cs_t_all} and \ref{fig:conc_th_t_all} is that the overall extent of entanglement in the oscillator-oscillator subsystem now decreases  with increase in $\bar{n}$.

\section{\label{sec:conclusion} Conclusion}
We have analysed a hybrid quantum system comprising  a pair of qubits and a pair of oscillators, described by the double Jaynes-Cummings model, and examined the effect of additional interactions such as the beamsplitter interaction between the oscillators,  and the dipole-dipole and Ising-type interactions between the qubits, on the entanglement dynamics. We find that the Ising type interaction  has significant impact on the dynamics, compared to the other forms of interaction. Our analysis is carried out for two types of entangled qubit states (Bell states). The oscillator initial states include pure states such as the ground states, Fock states and standard coherent states, as well as  thermal states. We have derived exact analytic expressions for concurrence for initial ground and higher Fock states oscillators and explained its dynamics, while for coherent state oscillators exact expressions for the state vectors are obtained. The role played by these initial states on the entanglement dynamics is investigated in detail. A useful result obtained here is that the extent to which the Ising interaction modifies the entanglement dynamics, is sensitive to the precise nature of the qubit initial state.  We have also carried out a detailed investigation on the important role of environmentally induced decoherence effects on the entanglement dynamics (see Appendix for details).  All the analysis that is carried out in this work and the results presented here assume perfect resonance between the qubits and the oscillators (detuning $\Delta = \omega-\omega_0 =0$). However, even with small detuning we have verified that these effects (ESD and entanglement transfer) persists, although the extent of it is little less (Appendix). These inferences corroborate very well the results presented in Ref.~\cite{Chan_2009}, where  ESD and entanglement transfer are shown to be prevented using off-resonant interactions.
In the field of entangled quantum information networks, it has been well documented~\cite{Shor_PRA_1995,Steane_PRL_1996,Jaeger_2014,Aolita_2015,Im_photonics_2022}  that  quantum error correction and quantum key distribution technology in particular,  allow for the possibility to restore even the weakest amount of degraded entanglement to full usefulness~\cite{Shor_PRA_1995,Steane_PRL_1996,Jaeger_2014,Im_photonics_2022} through various entanglement purification protocols, such as  entanglement distillation~\cite{pan1,Bennett_PRL_1996,Yamamoto_nature_2003} and entanglement concentration~\cite{Bennett_PRA_1996,Zhao_PRL_2003}. Unfortunately, such protocols fail completely when the  entanglement becomes strictly zero, i.e.,  during ESD. So any physical process that allows for the possibility of avoiding ESD completely would be of considerable interest.
Our investigation provides pointers towards these important aspects of entanglement dynamics, such as how ESD can be avoided completely, its onset can be delayed, its duration can be shortened etc., even in the presence of the environment, when different additional couplings are added to the basic DJC Hamiltonian.


\begin{acknowledgments}
I thank S.~Lakshmibala for suggestions during the preparation of the manuscript. 
\end{acknowledgments}

\appendix
\section{Concurrence for the DJC model with ground state oscillators}
\label{sec:DJC_appendix}
In this Section, we give some of the key steps leading to the exact analytical expressions for the qubit-qubit and oscillator-oscillator concurrences  for the DJC system (Eq.~(5)) in the simplest case when both the oscillators are initially in the ground states, excluding any additional interactions (see Ref.~\cite{Y_na__2006} for details). 

In the main text, we wrote down the state vectors $\ket{\Psi_{1}(t)}$ and $\ket{\Psi_{2}(t)}$  (Eq.~(13) and Eq.~(14)) for the two initial entangled qubit states $\ket{\psi_{1}(0)}$ and $\ket{\psi_{2}(0)}$ (Eq.~(8) and Eq.~(9)).
The time dependent coefficients  (obtained after solving the Schr\"odinger equation)  for $\ket{\psi_{1}(0)}$ are given by
\begin{subequations}
\begin{align}
 \label{eqn:psi1x1t}
 x_{1} &=   \cos\phi \, \cos\tau ,\\
 \label{eqn:psi1x2t}
 x_{2} &=   \sin\phi \,  \cos\tau  ,\\
 \label{eqn:psi1x3t}
 x_{3} &=  -i\cos\phi \, \sin\tau, \\
 \label{eqn:psi1x4t}
 x_{4} &=  -i\sin\phi \,  \sin\tau ,
\end{align} 
\label{eqn:xt}
\end{subequations}
while the coefficients for $\ket{\psi_{2}(0)}$ are 
\begin{subequations}
\begin{align}
 \label{eqn:psi1y1t}
 y_{1} &=   \cos\phi \, \cos^2\tau \, e^{-i\tilde{\omega} \tau},\\
 \label{eqn:psi1y2t}
 y_{2} &= - \cos\phi \, \sin^2\tau \, e^{-i\tilde{\omega} \tau},\\
 \label{eqn:psi1y3t}
 y_{3} &=  y_{4} = -i \cos\phi \, \sin\tau \cos\tau \, e^{-i\tilde{\omega} \tau},\\
 \label{eqn:psi1y5t}
 y_{5} &=   \sin\phi \,  \, e^{i\tilde{\omega} \tau},
\end{align} 
\label{eqn:yt}
\end{subequations}
 where $\tilde{\omega} = \omega/g_{\text{JC}}$.

The reduced density matrix $\rho_{1}(\tau)$ for the two qubits (obtained from Eq.~(13) by tracing out the oscillator contributions) can be explicitly written down in the basis $\ket{e\,e}$, $\ket{e\,g}$, $\ket{g\,e}$, $\ket{g\,g}$, and is given by
\begin{equation}
\rho_{1} = 
\begin{pmatrix}
 0 & & 0 & 0 & 0 \\
 0 & & \vert x_1\vert^2 & x_1 x_2^* & 0 \\
 0 & & x_1^* x_2 & \vert x_2\vert^2 & 0 \\
 0 & & 0 & 0 & \vert x_3\vert^2 + \vert x_4\vert^2 \\
\end{pmatrix}.
\label{eqn:rho1}
\end{equation}

Similarly,  $\rho_{2}(\tau)$, obtained from Eq.~(14), can be  written in the basis $\ket{e\,g}$, $\ket{e\,e}$, $\ket{g\,g}$, $\ket{g\,e}$ as 
\begin{equation}
\rho_{2} = 
\begin{pmatrix}
 \vert y_3\vert^2 & & 0 & 0 & 0 \\
 0 & & \vert y_1\vert^2 & y_1 y_5^* & 0 \\
 0 & & y_1^* y_5 & \vert y_2\vert^2 + \vert y_5\vert^2 & 0 \\
 0 & & 0 & 0 & \vert y_4\vert^2 \\
\end{pmatrix}.
\label{eqn:rho2}
\end{equation}
Now, the concurrence $\mathcal{C}$ (see, Eq.~(12) for definition) for any two-qubit density matrix  of the form
\begin{equation}
\rho = 
\begin{pmatrix}
 \rho_{11} & & 0 & 0 & 0 \\
 0 & & \rho_{22} & \rho_{23} & 0 \\
 0 & & \rho_{32} & \rho_{33} & 0 \\
 0 & & 0 & 0 & \rho_{44} \\
\end{pmatrix}.
\label{eqn:twoq_dm}
\end{equation}
has been shown~\cite{Eberly_PRL_2006} to be 
\begin{equation}
  \mathcal{C} = 2\,\text{max}\{0,\,\vert \rho_{23}\vert - \sqrt{\rho_{11}\rho_{44}}\,\}.
  \label{eqn:conc_defn}
\end{equation}
Comparing this with Eq.~\eqref{eqn:rho1} and Eq.~\eqref{eqn:rho2} we obtain the expressions for the qubit-qubit concurrences $ {\mathcal{C}_{\text{q}}}_1$ and $ {\mathcal{C}_{\text{q}}}_2$ (Eq.~(15) and Eq.~(17), respectively). Following similar procedure, we  obtain the equivalent expressions for the oscillator-oscillator concurrences shown in Eq.~(16) and Eq.~(18), respectively.

\section{DJC plus additional interaction Hamiltonians with ground state oscillators}
\label{sec:CDE_appendix}
 In this Section, we give details of the state vectors, each obtained after including additional interaction to the basic DJC Hamiltonian. 

For qubit-qubit entangled initial state $\ket{\psi_1(0)}$  in Eq.~(8), the state vectors corresponding to all the three cases have the same form given by Eq.~(13).
Now, the coefficients for an additional BS interaction take the form
\begin{subequations}
 \label{eqn:psi1xtbs}
\begin{align}
 \label{eqn:psi1x1tbs}
 x_{1} &= \cos\phi\, g(\tau_1,\tau_2)  - i \sin\phi\, h(\tau_1,\tau_2)\,, \\ 
 \label{eqn:psi1x2tbs}
 x_{2} &= \sin\phi \, g(\tau_1,\tau_2) - i \cos\phi\, h(\tau_1,\tau_2)\,, \\ 
 \label{eqn:psi1x3tbs}
 x_{3} &= -\frac{\tau}{\tau_2}\,  \Big(\sin\phi \, \sin\tau_1 + i \cos\phi \, \cos\tau_1 \Big) \sin\tau_2\,, \\
 \label{eqn:psi1x4tbs}
 x_{4} &= -\frac{\tau}{\tau_2}\,\Big(\cos\phi \, \sin\tau_1 + i \sin\phi \, \cos\tau_1 \Big) \sin\tau_2\,,
\end{align} 
\label{eqn:xtbs}
\end{subequations}
where $g(\tau_1,\tau_2)$ and $h(\tau_1,\tau_2)$ are defined in Eq.~(20). 

The time-dependent coefficients for an additional DD interaction have the same functional forms as in Eq.~\eqref{eqn:psi1xtbs} with the identification that $r_\textrm{b}=r_\textrm{d}$ and with $g(\tau_1,\tau_2)$ and $h(\tau_1,\tau_2)$ that are now defined in Eq.~(22).

The coefficients for an additional Ising interaction Hamiltonian are given by
\begin{subequations}
\begin{align}
 \label{eqn:psi1x1tis}
 x_{1} &= \cos\phi \left(\cos\tau_1+ i\sqrt{1-\frac{\tau^2}{\tau_1^2}}\, \sin\tau_1\right), \\
\label{eqn:psi1x2tis}
 x_{2} &=\sin\phi \left(\cos \tau_1+ i\sqrt{1-\frac{\tau^2}{\tau_1^2}}\, \sin \tau_1\right), \\
 \label{eqn:psi1x3tis}
 x_{3} &=- i\frac{\tau}{\tau_1} \cos\phi \sin\tau_1, \\
 \label{eqn:psi1x4tis}
 x_{4} &=-i \frac{\tau}{\tau_1} \sin\phi \sin\tau_1\,,
\end{align} 
\label{eqn:xtis}
\end{subequations}
where $\tau_1 = \sqrt{1+r_{\text{i}}^2}\, \tau$.

We, now, consider the other qubit-qubit entangled initial state $\ket{\psi_2(0)}$ (Eq.~(9)). The state vector after including an additional BS or dipole-dipole interaction is given in Eq.~(25). The time dependent coefficients for the state vector for the additional BS interaction are obtained using {\it Mathematica}. These are
\begin{subequations}
\begin{align}
 \label{eqn:psi2y1tbs1_new}
 y_{1} &=  \frac{4}{D_1}\left(A_1 \cos\tau_{+} + B_1 \cos\tau_{-} + C_1(r_b^2-1)\right)e^{-i\tilde{\omega} \tau} \cos\phi,\\
 \label{eqn:psi2y2tbs1_new}
 y_{2} &= \frac{4}{D_1}\left(A_2 \sin\tau_{+} - B_2 \sin\tau_{-} +C_1\right)e^{-i\tilde{\omega} \tau} \cos\phi ,\\
 \label{eqn:psi2y3tbs1_new}
 y_{3} &=  y_{4} = -i\frac{4}{D_3}\left(A_3 \sin\tau_{+} + B_3 \sin\tau_{-}\right) e^{-i\tilde{\omega} \tau} \cos\phi  ,\\
 \label{eqn:psi2y5tbs1_new}
 y_{5} &=  e^{i\tilde{\omega} \tau} \, \sin\phi ,\\
 \label{eqn:psi2y6tbs1_new}
  y_{6} &= y_{7} = i r_b \frac{N_6}{D_3} \left(\delta_+\, \sin\tau_{-} - \delta_{-}\, \sin \tau_{+}\right)  e^{-i\tilde{\omega} \tau} \cos\phi, \\
 \label{eqn:psi2y9tbs1_new}
 y_{8} &= y_9 = \frac{4r_b}{D_1}\left(A_8 \cos\tau_{+} + B_8 \cos\tau_{-} - C_1 \right) e^{-i\tilde{\omega} \tau} \cos\phi,
\end{align} 
\label{eqn:ytbs1_new}
\end{subequations}
where $\delta_\pm = \sqrt{\frac{1}{2}(5r_b^2+6\pm\Gamma_b)}$,  $\Gamma_b = \sqrt{9r_b^4 + 60r_b^2+4}$,$\tau_{\pm} = \delta_{\pm}\,t$, 
and
\begin{align}
    A_1 &= \left(r_b^4+2\right) \left(27 r_b^4+ 3(3 \Gamma + 16) r_b^2 + 2 (\Gamma +2)\right),\nonumber\\
    A_2 &= \left(r_b^4+2\right) \left(27 r_b^6+9 (\Gamma +15) r_b^4+15 (\Gamma +4) r_b^2+ 2 (\Gamma +2)\right), \nonumber\\
    A_3 &= \left(54 r_b^6+ 6\left(3 \Gamma+40\right)r_b^4 + 4\left(7 \Gamma +29\right)r_b^2  + 4(\Gamma + 2)\right)\delta_-,\nonumber\\
    A_8 &= \left(r_b^4+2\right) \left(27 r_b^4+(9 \Gamma +78) r_b^2 + 4 (\Gamma +2)\right),\nonumber\\
    B_1 &= 18 r_b^2 \Big(3 r_b^8+(\Gamma +28) r_b^6 + 2(3 \Gamma + 34) r_b^4 + 2(4 \Gamma +21) r_b^2 \nonumber \\
    &\qquad\qquad\qquad+2 (\Gamma +2)\Big),\nonumber\\
    B_2 &= 6 r_b^2 \Big(9 r_b^8 + 3\left(\Gamma +20\right)r_b^6 + 2(5 \Gamma+31)r_b^4 +2 (\Gamma -7) r_b^2\nonumber \\
    &\qquad\qquad\qquad- 2 (\Gamma +2)\Big),\nonumber\\
    B_3 &= 9 r_b^2 \left(3 r_b^6 +(\Gamma +22) r_b^4 + (4 \Gamma +30) r_b^2 +2 (\Gamma +2)\right)\delta_+,\nonumber\\
    B_8 &= 3 \Big(9 r_b^{10} + 3 (\Gamma +22) r_b^8+12 (\Gamma +5) r_b^6-4 (\Gamma +35) r_b^4 \nonumber \\
          &\qquad\qquad\qquad-20 (\Gamma +5) r_b^2 -4 (\Gamma +2) \Big), \nonumber \\
    C_1 &= \left(r_b^2-1\right) \Big(27 r_b^8 + 9 (\Gamma +28) r_b^6 + 6 (9 \Gamma +85) r_b^4 \nonumber \\
    &\qquad\qquad\qquad+2(23 \Gamma +76) r_b^2 + 4 (\Gamma +2)\Big),\nonumber \\
    D_1 &= \Gamma  \left(r_b^4+2\right) \left(9 r_b^4+3 (\Gamma +12) r_b^2+2 (\Gamma +2)\right)\delta _+^2, \nonumber 
\end{align}
\begin{align}
    D_3 &= \Gamma \left(9 r_b^4+3 (\Gamma +12) r_b^2 + 2 (\Gamma +2)\right) \delta _+^3 \delta _-,\nonumber \\
    N_6 &= 12 \sqrt{2}  \left(9 r_b^6 + 3(\Gamma+18) r_b^4 + 2  (4\Gamma + 23) r_b^2  +  2(\Gamma+ 2)\right).\nonumber
\end{align}

Following similar approach for the case of an additional dipole-dipole interaction, we get
\begin{subequations}
\begin{align}
 \label{eqn:psi2y1tdd1_new}
 y_{1} &= \left(A_{1_+} \cos\tau_+ - A_{1_-}\cos\tau_{-} +\frac{1}{2} \right) e^{-i\tilde{\omega} \tau} \cos\phi,\\
 \label{eqn:psi2y2tdd1_new}
 y_{2} &= \left(A_{2_+} \cos\tau_+ - A_{2_-}\cos\tau_{-} - \frac{1}{2} \right) e^{-i\tilde{\omega} \tau} \cos\phi ,\\
 \label{eqn:psi2y3tdd1_new}
 y_{3} &=  y_{4} = \frac{N_3}{D_3} \left(A_3\sin\tau_+ + B_3\sin\tau_-\right) e^{-i\tilde{\omega} \tau} \cos\phi  ,\\
 \label{eqn:psi2y5tdd1_new}
 y_{5} &=  e^{i\tilde{\omega} \tau} \, \sin\phi ,\\
 \label{eqn:psi2y6tdd1_new}
  y_{6} &= y_{7} = i \sqrt{2} \frac{r_d}{\Gamma }  \left(\frac{\sin \tau_+}{\delta _+}-\frac{\sin \tau_-}{\delta _-}\right) e^{-i\tilde{\omega} \tau} \cos\phi, \\
 \label{eqn:psi2y9tdd1_new}
 y_{8} &= y_9 = \frac{r_d}{\Gamma} \left(\cos\tau_+ - \cos\tau _- \right) e^{-i\tilde{\omega} \tau} \cos\phi,
\end{align} 
\label{eqn:ytdd1_new}
\end{subequations}
where $\delta _\pm=\sqrt{\frac{1}{2}(r_d^2+6\pm\Gamma_d)}$,  $\Gamma_d=\sqrt{r_d^4+12 r_d^2+4}$, $\tau_\pm = \delta_\pm t$, and
\begin{align}
    A_{1\pm} &= \frac{1}{4 \Gamma }\left(r_d^2+2\pm\Gamma\right),\nonumber \\
    A_{2\pm} &= \mp \frac{1}{4 \Gamma }\left(r_d^2-2\mp\Gamma\right),\nonumber \\
    N_3 &= -i \left(r_d^4 +(\Gamma +12) r_d^2+ 2 (\Gamma +2)\right),\nonumber \\
    D_3 &= 2 \Gamma ^3  \left(r_d^4+(\Gamma +8) r_d^2 + 2(\Gamma +2)\right) \delta _+ \delta _-,\nonumber \\
    A_3 &= \delta _- \left(r_d^6+(\Gamma +14) r_d^4 + 4 (2 \Gamma +7) r_d^2 + 4 (\Gamma +2)\right),\nonumber \\
    B_3 &= 4 \Gamma  \delta _+ r_d^2.\nonumber
\end{align}
It is straightforward to verify that setting $r_b=0$ ($r_d=0$) in Eq.~\eqref{eqn:ytbs1_new}  (Eq.~\eqref{eqn:ytdd1_new}) we get back Eq.~\eqref{eqn:yt}, which serves as a consistency check.

The two-qubit density matrix has the structure of an $X$ state
\begin{equation}
\rho_X = 
\begin{pmatrix}
 \rho_{11} & & 0 & 0 & \rho_{14} \\
 0 & & \rho_{22} & \rho_{23} & 0 \\
 0 & & \rho_{32} & \rho_{33} & 0 \\
 \rho_{41} & & 0 & 0 & \rho_{44} \\
\end{pmatrix},
\label{eqn:twoq_dm_x}
\end{equation}
where 
$\rho_{11} = \vert y_3\vert^2+\vert y_8\vert^2$, $\rho_{22} = \vert y_1\vert^2$, $\rho_{33} = \vert y_2\vert^2+\vert y_5\vert^2 + \vert y_6\vert^2+\vert y_7\vert^2$, $\rho_{44} = \vert y_4\vert^2+\vert y_9\vert^2$, $\rho_{14} = y_3^*y_9+y_4y_8^*$, $\rho_{23} = y_1^*y_5$. Now, for an $X$ state in Eq.~\eqref{eqn:twoq_dm_x}, the concurrence is defined as~\cite{eberly_qic}
\begin{equation}
    \mathcal{C} = 2\,\text{max}\left\{0,\, \vert \rho_{23}\vert - \sqrt{\rho_{11}\rho_{44}},\,\vert \rho_{14}\vert - \sqrt{\rho_{22}\rho_{33}}\right\}. 
\end{equation}

 In the presence of additional Ising interaction to the DJC Hamiltonian, the equations for the state vector remains simpler, as in Eq.~(14), but the coefficients get modified to
 \begin{subequations}
\begin{align}
 \label{eqn:psi1y1tis}
 y_{1} &= \tfrac{1}{2}\cos\phi \left(\cos\tau_2 - i\frac{\tau_1}{\tau_2}\sin\tau_2 + e^{-i \tau_1}\right)e^{-i\tilde{\omega}\tau} ,\\
 \label{eqn:psi1y2tis}
 y_{2} &= \tfrac{1}{2}\cos\phi \left(\cos\tau_2 - i\frac{\tau_1}{\tau_2}\sin\tau_2 - e^{-i \tau_1}\right)e^{-i\tilde{\omega}\tau}, \\
 \label{eqn:psi1y3tis}
 y_{3} &= -i \frac{\tau}{\tau_2} \cos\phi\sin \tau_2\,e^{-i\tilde{\omega}\tau},\\
 \label{eqn:psi1y4tis}
 y_{4} &= -i \frac{\tau}{\tau_2} \cos\phi \sin\tau_2\,e^{-i\tilde{\omega}\tau}, \\
 \label{eqn:psi1y5tis}
 y_{5} &=  \sin\phi\, e^{-i\tau_1} e^{i\tilde{\omega}\tau},
\end{align} 
\label{eqn:ytis}
\end{subequations}
 where $\tilde{\omega} = \omega/g_{\text{JC}}$,  $\tau_1 = r_{\text{i}} \tau$ and $\tau_2 = \sqrt{r_{\text{i}}^2+4} \, \tau$.

\section{DJC model with oscillator Fock states}
\label{sec:appendix_fock}

When both the oscillators are initially in arbitrary Fock states $\ket{n}$ and $\ket{m}$ respectively, the state vector corresponding to the qubit-qubit entangled initial state $ \ket{\psi_{1}(0)}$ is given by
\begin{align}
 \ket{\psi_{1}(\tau)} &= x_{1} \ket{e\, g\, n\, m} + x_{2} \ket{g\, e\, n\, m} + x_{3} \ket{g\, g\, n+1\, m}  \nonumber \\[2pt]
                      &+ x_{4} \ket{g\, g\, n\, m+1} + x_{5} \ket{e\, e\, n-1\, m} \nonumber \\[2pt]
                      &+ x_{6} \ket{e\, e\, n\, m-1} + x_{7} \ket{g\, e\, n+1\, m-1} \nonumber \\[2pt]
                      &+ x_{8} \ket{e\, g\, n-1\, m+1}.
 \label{eqn:psi1t_fock}
\end{align}
Note that $x_5$ and $x_8$ ($x_6$ and $x_7$) vanish if $n=0$ ($m=0$). Solving the Schr\"odinger equation we get
\begin{subequations}
 \label{eqn:psi1xvec_fock}
\begin{align}
 \label{eqn:psi1x1tf}
    x_1 &= \cos\phi \cos\tau_1 \cos\tau_4 \,e^{-i\tilde{\omega}_{nm}\tau}, \\[2pt]
 \label{eqn:psi1x2tf}
    x_2 &= \sin\phi \cos\tau_2 \cos \tau_3 \,e^{-i\tilde{\omega}_{nm}\tau}, \\[2pt]
 \label{eqn:psi1x3tf}
    x_3 &= -i \cos\phi \sin\tau_1 \cos\tau_4 \,e^{-i\tilde{\omega}_{nm}\tau}, \\[2pt]
 \label{eqn:psi1x4tf}
    x_4 &= -i \sin\phi \sin\tau_2 \cos\tau_3 \,e^{-i\tilde{\omega}_{nm}\tau}, \\[2pt]
 \label{eqn:psi1x5tf}
    x_5 &= -i \sin\phi \cos\tau_2 \sin\tau_3 \,e^{-i\tilde{\omega}_{nm}\tau}, \\[2pt]
 \label{eqn:psi1x6tf}
    x_6 &= -i \cos\phi \cos\tau_1 \sin\tau_4 \,e^{-i\tilde{\omega}_{nm}\tau}, \\[2pt]
 \label{eqn:psi1x7tf}
    x_7 &= -\cos\phi \sin\tau_1 \sin\tau_4 \,e^{-i\tilde{\omega}_{nm}\tau}, \\[2pt]
 \label{eqn:psi1x8tf}
    x_8 &= -\sin\phi \sin\tau_2 \sin\tau_3 \,e^{-i\tilde{\omega}_{nm}\tau}.
\end{align}
\end{subequations}
where $\tilde{\omega}_{nm} = (n+m)\tilde{\omega}$.  $\tau_1 = \sqrt{n+1}\, \tau$, $\tau_2 = \sqrt{m+1}\, \tau$, $\tau_3 = \sqrt{n}\, \tau$ and $\tau_4 = \sqrt{m}\, \tau$.
Corresponding to the state vector in Eq.~\eqref{eqn:psi1t_fock}, the reduced two-qubit density matrix obtained in the basis $\ket{ee},\, \ket{eg},\, \ket{ge},\, \ket{gg}$ has the exact form as in Eq.~\eqref{eqn:twoq_dm}, where $\rho_{11} = \vert x_5\vert^2 + \vert x_6\vert^2$, $\rho_{22} = \vert x_1\vert^2 + \vert x_8\vert^2$, $\rho_{33} = \vert x_2\vert^2 + \vert x_7\vert^2$, $\rho_{44} = \vert x_3\vert^2 + \vert x_4\vert^2$ and $\rho_{23} =  x_1 x_2^*$.  Next, using Eq.~\eqref{eqn:conc_defn} we obtain the exact analytical expression for the concurrences which is given in Eq.~(27).

Similarly, corresponding to the other initial entangled state $ \ket{\psi_{2}(0)}$, the state at a subsequent time $\tau$ is given by
\begin{align}
 \ket{\psi_{2}(\tau)} &= y_{1}\ket{e\, e\, n\, m} + y_{2}\ket{g\, g\, n+1\, m+1} \nonumber \\[2pt]
                          &+ y_{3}\ket{e\, g\, n\, m+1} +  y_{4}\ket{g\, e\, n+1\, m} \nonumber \\[2pt]
                          &+ y_{5}\ket{g\, g\, n\, m}   + y_{6}\ket{e\, e\, n-1\, m-1}\nonumber \\[2pt]
                          &+ y_{7}\ket{e\, g\, n-1\, m} + y_{8}\ket{g\, e\, n\, m-1}.
\label{eqn:psi2t2} 
\end{align}

The time-dependent coefficients are given by
\begin{subequations}
 \label{eqn:psi1yvec_fock}
\begin{align}
 \label{eqn:psi1y1tf}
 y_{1} &=   \cos\phi \, \cos\tau_1  \cos\tau_2  \, e^{-\tilde{\omega}_{nm_{+}}\tau}, \\[2pt]
 \label{eqn:psi1y2tf}
 y_{2} &= - \cos\phi \, \sin\tau_1   \sin\tau_2  \, e^{-\tilde{\omega}_{nm_{+}}\tau},\\[2pt]
 \label{eqn:psi1y3tf}
 y_{3} &= -i \cos\phi \, \cos\tau_1 \sin\tau_2 \, e^{-\tilde{\omega}_{nm_{+}}\tau},\\[2pt]
 \label{eqn:psi1y4tf}
 y_{4} &= -i \cos\phi \, \sin\tau_1 \cos\tau_2 \, e^{-\tilde{\omega}_{nm_{+}}\tau},\\[2pt]
 \label{eqn:psi1y5tf}
 y_{5} &=  \sin\phi \,  \cos\tau_3 \cos\tau_4 \, e^{-\tilde{\omega}_{nm_{-}}\tau},\\[2pt]
 \label{eqn:psi1y6tf}
 y_{6} &=  -\sin\phi \,  \sin\tau_3  \sin\tau_4 \, e^{-\tilde{\omega}_{nm_{-}}\tau},\\[2pt]
 \label{eqn:psi1y7tf}
 y_{7} &= -i \sin\phi \,  \sin\tau_3  \cos\tau_4 \, e^{-\tilde{\omega}_{nm_{-}}\tau},\\[2pt]
 \label{eqn:psi1y8tf}
 y_{8} &= -i \sin\phi \,  \cos\tau_3 \sin\tau_4 \, e^{-\tilde{\omega}_{nm_{-}}\tau},
\end{align} 
\end{subequations}
where $\tilde{\omega}_{nm_{\pm}} = e^{-i(n+m\pm1)\tilde{\omega}\tau}$, $\tau_1 = \sqrt{n+1}\, \tau$, $\tau_2 = \sqrt{m+1}\, \tau$, $\tau_3 = \sqrt{n}\, \tau$ and $\tau_4 = \sqrt{m}\, \tau$.

Corresponding to the state vector in Eq.~\eqref{eqn:psi1t_fock}, the reduced two-qubit density matrix obtained in the basis $\ket{ee},\, \ket{eg},\, \ket{ge},\, \ket{gg}$ has the form as in Eq.~\eqref{eqn:twoq_dm}, where $\rho_{11} = \vert y_3\vert^2 + \vert y_7\vert^2$, $\rho_{22} = \vert y_1\vert^2 + \vert y_6\vert^2$, $\rho_{33} = \vert y_2\vert^2 + \vert y_5\vert^2$, $\rho_{44} = \vert y_4\vert^2 + \vert y_8\vert^2$ and $\rho_{23} =  y_1 y_5^*$. After some simple algebra we  obtain the equivalent expression for the concurrence (see Eq.~(28)).

\section{DJC model with coherent oscillator states}
\label{sec:appendix_cs}
Now consider the set of initial states where both the fields are in standard coherent states $\ket{\alpha}$.
Using the photon number expansions of $\ket{\alpha} = \sum_n q_{n}(\alpha) \ket{n}$, where $q_{n}(\alpha)  = e^{-\frac{1}{2}\vert\alpha\vert^{2}}\frac{\alpha^{n}}{\sqrt{n!}}$, it is now straightforward to write down the expressions for the state vectors for the two initial configurations. 
These are
\begin{align}
 \label{eqn:psi3t}
 \ket{\psi_{1}(\tau)} = &\sum_{n,m=0}^{\infty} q_{n}\, q_{m}\,\big\{
                      x_{1} \ket{e\, g\, n\, m}    + x_{2} \ket{g\, e\, n\, m}  \nonumber \\ 
                    &+ x_{3} \ket{g\, g\, n+1\,m}  + x_{4} \ket{g\, g\, n\, m+1} \nonumber \\ 
                    &+ x_{5} \ket{e\, e\, n-1\, m} + x_{6} \ket{e\, e\, n\, m-1}  \nonumber \\ 
                    &+ x_{7} \ket{g\, e\, n+1\, m-1} + x_{8} \ket{e\, g\, n-1\, m+1} \big\},\\
\label{eqn:psi4t} 
 \ket{\psi_{2}(\tau)} = &\sum_{n,m=0}^{\infty} q_{n}\, q_{m}\,\big\{
                         y_{1} \ket{e\, e\, n\, m} + y_{2}\ket{g\, g\, n+1\, m+1} \nonumber \\
                      &+ y_{3} \ket{e\, g\, n\, m+1} +  y_{4}\ket{g\, e\, n+1\, m} \nonumber \\ 
                      &+ y_{5} \ket{g\, g\, n\, m} + y_{6}\ket{e\, e\, n-1\, m-1}\nonumber \\
                      &+ y_{7} \ket{e\, g\, n-1\, m} + y_{8}\ket{g\, e\, n\, m-1}\big\},
\end{align}
respectively, where  $x_{i}(\tau)$ and $y_{i}(\tau)$ ($i=1,\, 2,\ldots, 8$) are given in Eq.~\eqref{eqn:psi1xvec_fock} and Eq.~\eqref{eqn:psi1yvec_fock}, respectively.
From the above two state vectors, we obtain the corresponding reduced qubit-qubit and oscillator-oscillator density matrices and then numerically compute the concurrence and LN following the procedure outlined above.

\section{Role of detuning}
\label{sec:appendix_detuning}
Thus far we have restricted our analysis to the zero detuning limit, i.e., we have assumed perfect frequency matching between the qubits and the oscillators. This is indeed an extremely idealised assumption to make. But this way it was easy to figure out that the effects imprinted on the dynamics were entirely due to the additional interactions, and not because of the imperfect matching of frequencies. However, in practice maintaining such perfect resonance is indeed a challenging task. 
\begin{figure}[ht!]
  \centering
  \includegraphics{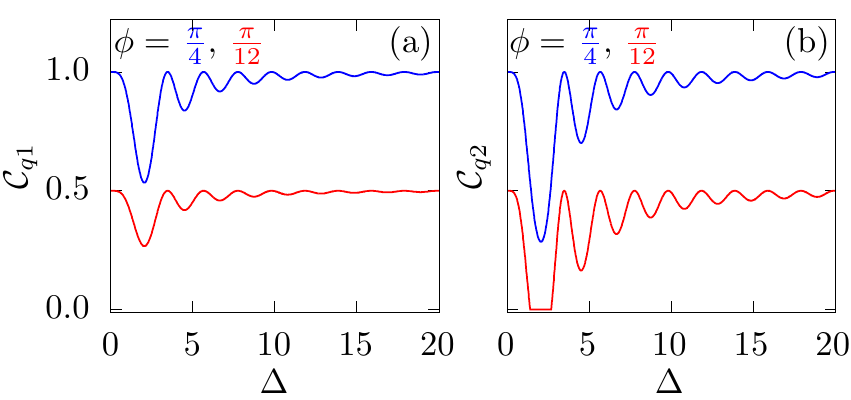}
  \caption{The role of imperfect matching of frequencies between the qubits and the oscillators on ${C_{\text{q}}}_1$ and ${C_{\text{q}}}_2$  for the standard DJC model. Panels (a) and (b) correspond to initial state $\ket{\psi_1(0)}$ and $\ket{\psi_2(0)}$, respectively with $\phi = \frac{\pi}{4}$ (blue) and $\frac{\pi}{12}$ (red). Initially, the oscillators are prepared in the ground states. In both panels, we have set the time for which maximum entanglement occurs. In the large detuning limit, we clearly see the concurrences for the respective cases saturating to the initial value. In other words, in the dispersive limit, the JC coupling between the qubit and the oscillator gets suppressed so that the dynamics of the system fails to change the initial entanglement.} 
  \label{fig1}
\end{figure}
In this Section, we analyse the role of imperfect matching of frequencies on the entanglement dynamics. To this end, let us first consider the standard DJC model (i.e., in the absence of any additional couplings). For simplicity, we assume equal frequencies $\omega_0$  ($\omega$) for both the qubits (oscillators). The detuning is denoted by $\Delta = \omega - \omega_0$. 

As before, the two qubits are initially prepared in either of the symmetric or antisymmetric  entangled Bell states ($\ket{\psi_1(0)}=\cos\phi \ket{e\,g} + \sin\phi \ket{g\,e}$ or $\ket{\psi_2(0)}=\cos\phi \ket{e\,e} + \sin\phi \ket{g\,g}$). We also assume that the oscillators are prepared in the ground states.

\begin{figure}[ht!]
  \centering
  \includegraphics{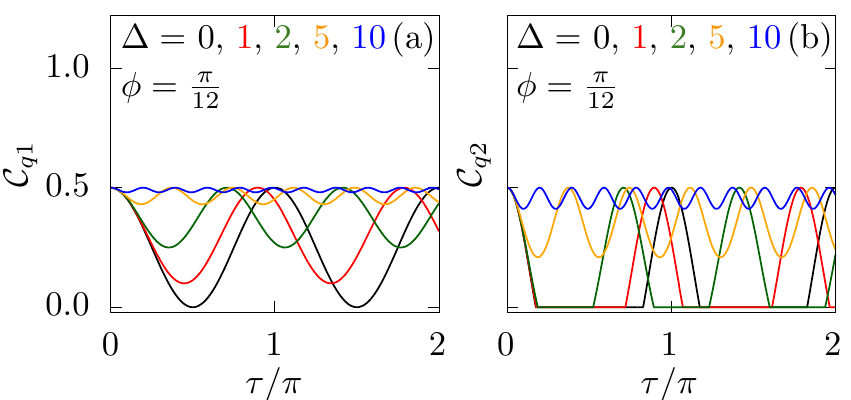}
  \caption{The time evolution of ${C_{\text{q}}}_1$ and ${C_{\text{q}}}_2$ for the DJC model with for different values of $\Delta$. Panels (a) and (b) correspond to initial state $\ket{\psi_1(0)}$ and $\ket{\psi_2(0)}$, respectively with $\phi = \frac{\pi}{12}$. Initially, the oscillators are prepared in the ground states.  As the detuning increases, the dynamics of ${C_{\text{q}}}_1$ and ${C_{\text{q}}}_2$  saturates to the initial value. } 
  \label{fig2}
\end{figure}

The expressions for the qubit-qubit concurrence for these two initial states are given by~\cite{Y_na__2006}
\begin{align}
    \mathcal{C}_{q1} &= \sin (2 \phi ) \left(1- 4N\sin^2\tau\right), \\
    \mathcal{C}_{q2} &= \max\left[0, \mathcal{C}_{q1} \left(1 - \frac{\left(1+\cos(2\phi)\right)4N\sin^2\tau)} {\sin(2\phi)} \right)\right],
\end{align}
where $\tau=\frac{1}{2} D t$, $D = \sqrt{\Delta ^2+4 g^2}$, $N = (g/D)^2$, and $g_{\text{JC}}^{(1)} = g_{\text{JC}}^{(2)}=g$.
In Figs.~\ref{fig1} and \ref{fig2}, we have plotted the qubit-qubit concurrences ${C_{\text{q}}}_1$ and ${C_{\text{q}}}_2$ for these two different initial configurations as a function of $\Delta$ and time, respectively. Both figures show that with small nonzero detuning the dynamics remain qualitatively  similar. However, the degree to which  the entanglement dynamics, more importantly,  ESD (see panel (b) in both figures) can be manipulated by introducing considerably large  off-resonant interactions are also depicted in Figs.~\ref{fig1} and \ref{fig2}. 
 In fact, by looking at  these figures, one can analyse the transition from the resonance dynamics to the dynamics of the `dispersive limit', where the detuning $\Delta$ is very very large (compared to the coupling strength $g$), resulting  in the suppression of the qubit-oscillator coupling.
 The physics of the dispersive limit is also easy to analyse analytically. It is well known that the  JC Hamiltonian involving a single qubit and a single oscillator $H_{\text{JC}}=\tfrac{1}{2}\omega_{q}  \sigma_{z} + \omega_o a^{\dagger}a + g \left(a \,\sigma_{+} + a^\dagger \sigma_-\right)$ in the large detuning limit, $\Delta = \omega_q-\omega_o \gg g$, (or equivalently,  $\frac{g}{\Delta}\ll 1$), can be written down as~\cite{Dodonov_2003}
\begin{equation}
H_{\textrm{eff}} = U H_{\text{JC}} U^\dagger = \left(\omega_{o}+\frac{g^2}{\Delta}\sigma_{z}\right)a^\dagger a + \frac{1}{2}\left(\omega_q + \frac{g^2}{\Delta}\right) \sigma_z,
\label{heff}
\end{equation}
where $U=\exp\left[\frac{g}{\Delta}\left(a \,\sigma_{+} + a^\dagger \sigma_-\right)\right]$.
Therefore, in the dispersive limit, the oscillator frequency is shifted by an amount $\pm \frac{g^2}{\Delta}$ (depending on the state of the qubit) whereas the qubit energy levels are shifted by an amount $\frac{g^2}{\Delta}$. 
Similar reasoning straightforwardly carries over to the double JC dynamics as well, and this results in Figs.~\ref{fig1} and \ref{fig2}. 
These results corroborate well with the inferences drawn in Ref.~\cite{Chan_JPhysB_2009}.

\begin{figure}[ht!]
  \centering
  \includegraphics{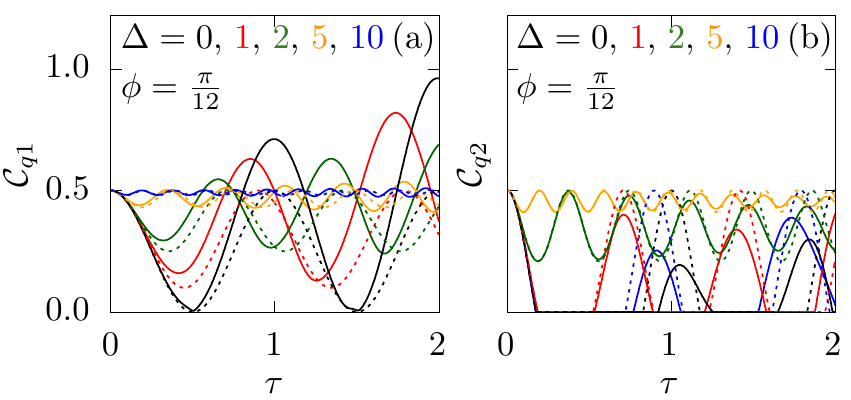}
  \caption{The time evolution of ${C_{\text{q}}}_1$ and ${C_{\text{q}}}_2$ for the DJC model with additional beamsplitter coupling between the oscillators for different values of $\Delta$.  The solid (dashed) lines correspond to $r_b=0.2$ (0). Panels (a) and (b) correspond to initial state $\ket{\psi_1(0)}$ and $\ket{\psi_2(0)}$, respectively with $\phi = \frac{\pi}{12}$. Initially, he oscillators are prepared in the ground states. } 
  \label{fig3}
\end{figure}

Now, with additional interactions, the inferences remain similar, as can be seen in Fig.~\ref{fig3} where the concurrence for the two different initial configurations are plotted for different values of $\Delta$ when additional beamsplitter coupling is considered.


 We mention in passing that apart from nonzero detuning another important aspect pertains to the changes in the entanglement dynamics observed when unequal coupling strengths are considered, i.e., when $g_{\text{JC}}^{(1)} \ne g_{\text{JC}}^{(2)}$. Results on this are also reported earlier~\cite{Chan_JPhysB_2009} where it is  shown that one can engineer the transfer process so that the initial entanglement is fully transferred into a desired qubit pair just by considering unequal coupling strengths. 

\section{Long time entanglement dynamics}
\label{sec:appendix_long_time}
\begin{figure}[ht!]
  \centering
  \includegraphics{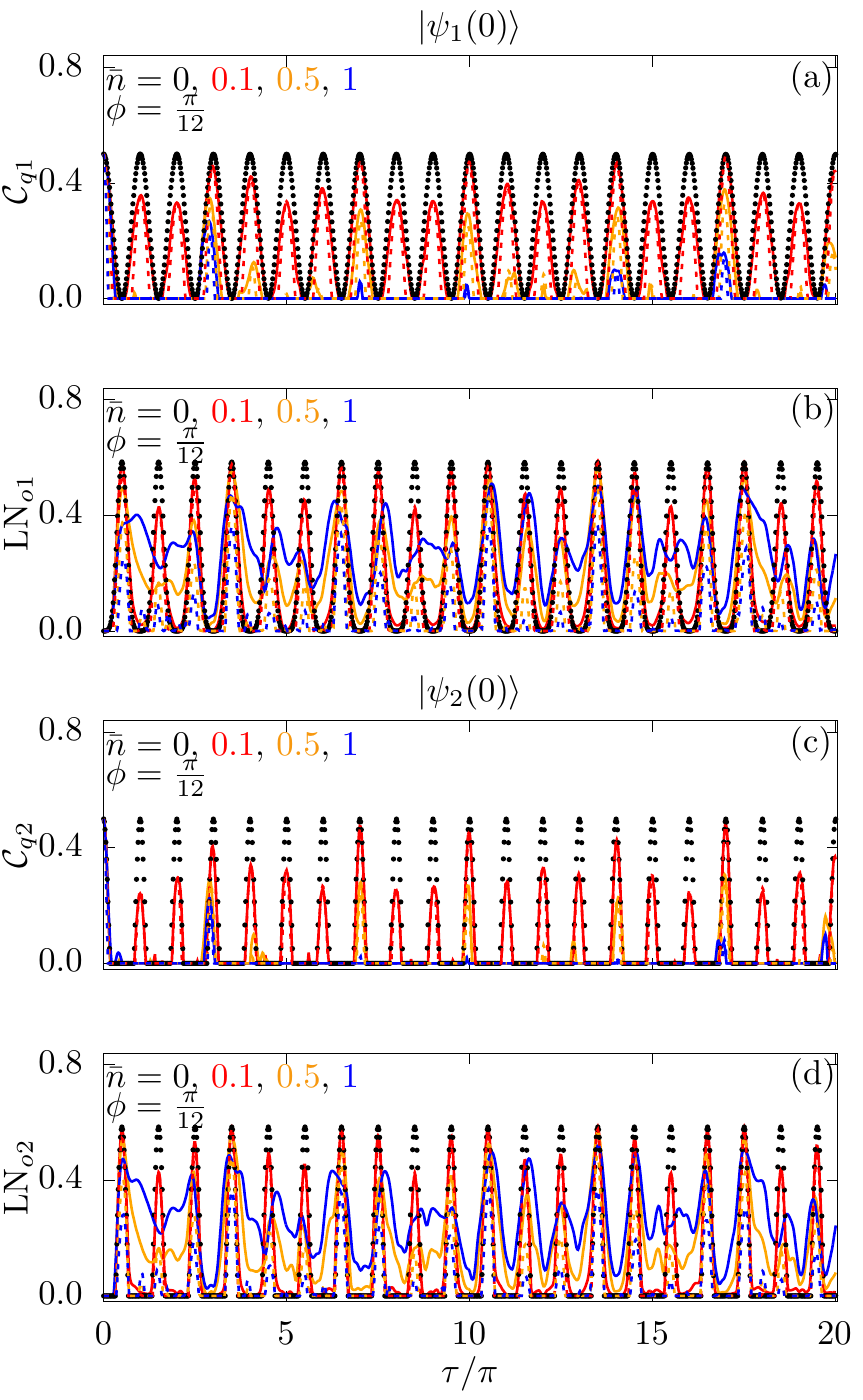}
  \caption{The long time evolution of qubit-qubit concurrence and oscillator-oscillator logarithmic negativity (LN)  for the double JC model with $\phi = \frac{\pi}{12}$. Both the oscillators are initially prepared either in the coherent states (solid lines) or in thermal states (dashed lines) with $\bar{n}=0,\, 0.1$, 0.5 and 1 respectively. The dotted lines correspond to the ground state oscillators. Although during the short time the dynamics tends  to exhibit decaying behavior on top of a oscillatory dynamics, this oscillatory dynamics continues at large times (i.e., it does not  converge to any specific value) without settling to a specific value.} 
  \label{fig4}
\end{figure}

So far we have focused only on the early time entanglement dynamics.
In this Section, we  analyse the long time dynamics of the qubit-qubit and oscillator-oscillator entanglement. For the standard DJC model Hamiltonian, the long-time entanglement dynamics is depicted in Fig.~\ref{fig4}, for both initial coherent and thermal oscillator states. As always, the qubits are initially prepared in either of the two maximally entangled Bell states $\ket{\psi_1(0)}$ or  $\ket{\psi_2(0)}$. The long time dynamics continue to exhibit somewhat complex imperfect oscillatory dynamics without converging to any specific value. Similar inferences can be drawn even when additional couplings are added to the standard DJC Hamiltonian. 

Another important thing to note is that all these inferences hold because we have neglected the (inevitable) system-environment interactions. In that case, of course, there will be (exponential) decay of entanglement and the entanglement will eventually vanish (see the Section on environmentally induced effects below).

\section{Experimental realisation of various system Hamiltonians}
 \label{sec:appendix_experiment}
The DJC Hamiltonian and its various generalisations considered in the article can, in principle, be realised in various quantum platforms, such as trapped ions, superconducting qubits and solid state system. In the following, we present a very brief summary of such realizations in the context of trapped ions. 

Consider two (say, $^{40}$Ca$^+$) ions trapped in a linear Paul trap (see, Refs.~\cite{wineland, lemmer_quantum_2018} for details). The degrees of freedom associated with the motion of the ions can effectively be approximated as quantum harmonic oscillators. On the other hand, ions' internal/electronic degrees of freedom, under the two-level approximation, mimic the role of qubits. For instance, in the case of $^{40}$Ca$^+$ ions, the two levels of the (optical) qubit are $\ket{g} \rightarrow 4\, {}^{2}S_{1/2}$ (electronic ground state),
$\ket{e} \rightarrow 4\, {}^{2}D_{5/2}$ (metastable  excited state with lifetime $T_1 \approx 1.2$s). The transition frequency $\omega_0$ is in the optical regime ($729$nm).
In the Lamb-Dicke regime, the interaction between each ion's motional degree of freedom (oscillator) and the internal degree of freedom (qubit),  under the rotating wave approximation would lead to JC interaction (when the laser is appropriately tuned at the first red sideband transition). The coupling coefficient $g_\text{JC} \equiv \frac{1}{2}\eta\Omega_0$, where $\eta$ is the Lamb-Dicke parameter and $\Omega_0$ is the resonance Rabi frequency. Therefore, with two ions in the trap, it is possible to realise the standard DJC Hamiltonian.

The DJC plus the beamsplitter Hamiltonian, on the other hand, is a special case of a more generic model, popularly known as the Jaynes-Cummings-Hubbard (JCH) model which involves $N$ bosonic modes and $N$ two-level systems (see, Refs.~\cite{Ivanov_PRA_2009,Toyoda_PRL_2013,Silpa_PRA_2021,silpa_2022} for details where these type of Hamiltonians are experimentally realised using trapped ions). In a trap where $N$ ions are trapped, apart from the DJC interaction, the beamsplitter interaction $\sum_{j<k}^{N} \frac{\kappa_{jk}}{2}(a_j a_k^\dagger + a_j^\dagger a_k)$ is realised by the phonon hopping interactions between motional modes of the $i$th and the $j$th ion.
$\kappa_{ij} \equiv e^2/{4\pi\epsilon_0 m d_{ij}^3\omega_i}$ is the hopping rate between  the $i$th and $j$th ion sites, where $m$ is the mass of he ion and $d_{ij}$ is the distance between the $i$th and the $j$th ion. The phonon hopping happens due to the Coulomb interaction, which can be controlled  easily by adjusting $d_{ij}$~\cite{Ivanov_PRA_2009}. Once again, with two ions in the trap, our model Hamiltonian $H_{\text{JDC}}+  \frac{\kappa_{12}}{2}(a_1 a_2^\dagger + a_2^\dagger a_1)$ can be easily realised experimentally. In particular, see Refs.~\cite{Toyoda_PRL_2013,Silpa_PRA_2021} where exactly this model Hamiltonian is simulated experimentally with two ions.

Moving on, the spin-spin interactions, both the dipole-dipole and the Ising types, have also been implemented in various ion trap set ups over the years (for details regarding different possible implementations, see, for instance, Refs.~\cite{Peng_PRA_1993,Porras_PRL_2004,Senko_PRX_2015,Wall_PRA_2017,Campbell_PRL_2020}). In the case of coherently trapped atoms, the dipole-dipole term describes the effect of the virtual off-resonance photon exchange between such atoms  which has to be included in addition to the JC interaction~\cite{Peng_PRA_1993}. 
Further work shows that when two atoms are trapped in microcavities that interact via the exchange of virtual photons, both atoms experience a Stark shift that depends on the state of the partner atom~\cite{Hartmann_PRL_2007}. This conditional Stark shift eventually leads to an effective Ising interaction between them.

\section{Environmental induced decoherence effects}
\label{sec:appendix_decoh}
Although hybrid quantum systems are largely robust against dissipative losses, still their role on the entanglement dynamics cannot be discarded entirely. 
The typical sources of losses that we examine here are the  dissipation and dephasing of the oscillators and the qubits, as they interact with the environment. We assume, for simplicity, that the oscillators and the qubits effectively interact with the same reservoir with temperature $\bar{n}_{\text{th}}$. 
We  ignore the Lamb shift contribution  leading to a small renormalization of the system energy levels. If $\rho_{S}(\tau)$ denotes the reduced system density matrix (obtained after tracing out the environmental degrees of freedom from the joint system-environment density matrix) at a later time $\tau$, the  master equation in the Lindblad formalism is given by
\begin{equation}
 \frac{d}{d\tau} \rho_{S}(\tau) = -i \left[H, \, \rho_{S}(\tau)\right] + \mathcal{L}[\rho_{S}(\tau)]
\end{equation}
\begin{figure}[ht!]
  \centering
  \includegraphics{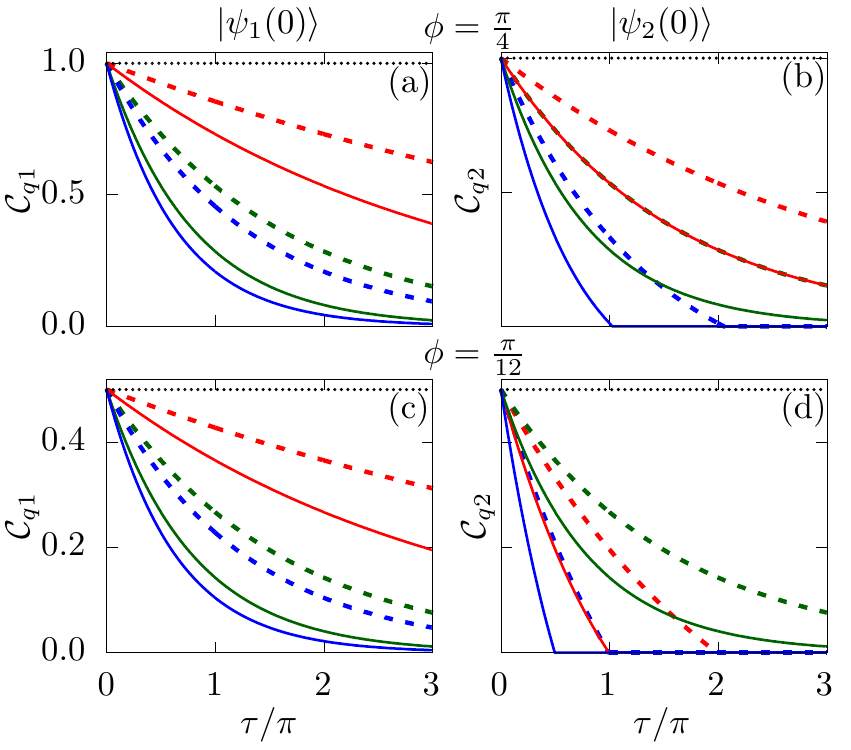}
  \caption{Stability/robustness of initially prepared entanglement under the influence of vacuum fluctuation, i.e, when $\bar{n}_{\text{th}}=0$:  ${C_{\text{q}}}_1$ (left) and ${C_{\text{q}}}_2$ (right) corresponding to initial entangled Bell state $\ket{\psi_1(0)}$ and $\ket{\psi_2(0)}$, respectively are displayed for $\phi = \frac{\pi}{4}$ (top) and $\frac{\pi}{12}$ (bottom). The red, green and the blue curves correspond to the loss of entanglement due to dissipation (i.e., $\lambda_d = 0$), dephasing (i.e., $\lambda_r = 0$), and both dissipation and dephasing respectively. The dashed (solid) curves correspond to the decay rates equal to $0.05$ (0.1). The black dotted lines correspond to initial entanglement. } 
  \label{fig_stability}
\end{figure}
where the dissipator $\mathcal{L}$ has the usual form
\begin{equation}
  \mathcal{L}[\rho_{S}] = \sum_{k} \left( L_{k} \rho_{S} L_{k}^{\dagger} -  \frac{1}{2}\left[\rho_{S},\, L_{k}^{\dagger} L_{k}\right] \right).
\end{equation}
Here, $L_{k} = \sqrt{\lambda}_{k} A_{k}$ are the  standard Lindblad superoperators for dissipation, and the environment couples to the system through the operators $A_{k}$ with coupling rates $\lambda_{k}$. The Lindblad operators for the oscillators are $a$, $a^{\dagger}$, $b$, $b^{\dagger}$ with decay rates $\sqrt{\lambda_{a}(1+\bar{n}_{\text{th}})}$, $\sqrt{\lambda_{a}\, \bar{n}_{\text{th}}}$, $\sqrt{\lambda_{b}(1+\bar{n}_{\text{th}})}$ and $\sqrt{\lambda_{b}\, \bar{n}_{\text{th}}}$, respectively. Similarly for the qubits, the corresponding operators for dissipation are $\sigma^{(i)}_{-}$ and $\sigma^{(i)}_{+}$  with respective decay rates $\sqrt{\lambda_{r}^{(i)}(1+\bar{n}_{\text{th}})}$, $\sqrt{\lambda_{r}^{(i)}\, \bar{n}_{\text{th}}}$  ($i=1,\,2$). For ease of analysis, we assume that $\lambda_a = \lambda_b =  \lambda_r^{(1)} = \lambda_r^{(2)} = \lambda_r$. On the other hand, for the dephasing we have the corresponding Lindblad operators $a^\dagger a$, $b^\dagger b$ and  $\sigma^{(i)}_{z}$. Once again, for simplicity we assume  equal dephasing rate $\sqrt{\lambda_{d}}$.  For the environment, we consider two specific cases: a vacuum or a zero temperature bath, ($\bar{n}_{\text{th}}=0$) and a thermal bath with nonzero $\bar{n}_{\text{th}}$.

In  the former case  ($\bar{n}_{\text{th}}=0$), the system  decoheres through vacuum fluctuations. Thus, only the Lindblad operators  $a$ and $b$ contribute, leading to spontaneous decays. Similarly, for the qubits, only $\sigma^{(i)}_{-}$ ($i=1,\,2$) contribute for the dissipation. The dephasing is  anyway independent of the bath temperature.  It is in general difficult to get exact analytical expressions for the state vectors in the presence of system-environment interactions.  Therefore, we rely on numerical observations for inferences. We have used the {\it qutip} library~\cite{qutip} for numerical analysis.

\begin{figure}[ht!]
  \centering
  \includegraphics{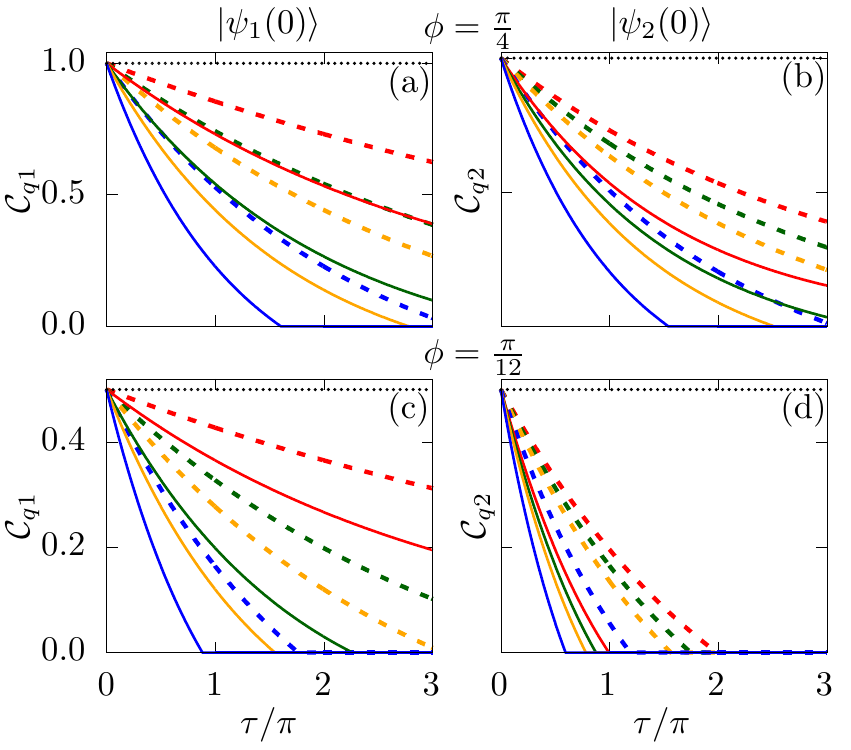}
  \caption{Stability/robustness of initially prepared entanglement under the influence of thermal bath with $\bar{n}_{\text{th}}=0$ (blue), 0.1 (green), 0.2 (yellow), 0.5 (blue):  ${C_{\text{q}}}_1$ (left) and ${C_{\text{q}}}_2$ (right) corresponding to initial entangled Bell state $\ket{\psi_1(0)}$ and $\ket{\psi_2(0)}$, respectively are displayed for $\phi = \frac{\pi}{4}$ (top) and $\frac{\pi}{12}$ (bottom). The dashed (solid) curves correspond to $\lambda_r=0.05$ (0.1). The black dotted lines correspond to initial entanglement. } 
  \label{fig_stability_nth}
\end{figure}

Before we let the full system interact with the environment let us first examine the stability/robustness of the initially prepared qubit-qubit pairwise entanglement~\cite{Cai_PRA_2005} when the qubits interact with the environment. 
In Fig.~\ref{fig_stability}, we have displayed this stability of the two entangled Bell states under the influence of vacuum fluctuation ($\bar{n}{\text{th}}=0$). Two important points can be identified immediately: (i) for any values of $\phi$, $\ket{\psi_{1}(0)}$ is more stable than $\ket{\psi_{2}(0)}$ under environmental induced losses, and (ii) for both the initial entangled states, maximally entangled Bell states ($\phi=\frac{\pi}{4}$) fare better than partially entangled Bell states.

\begin{figure*}[ht!]
  \centering
    \includegraphics{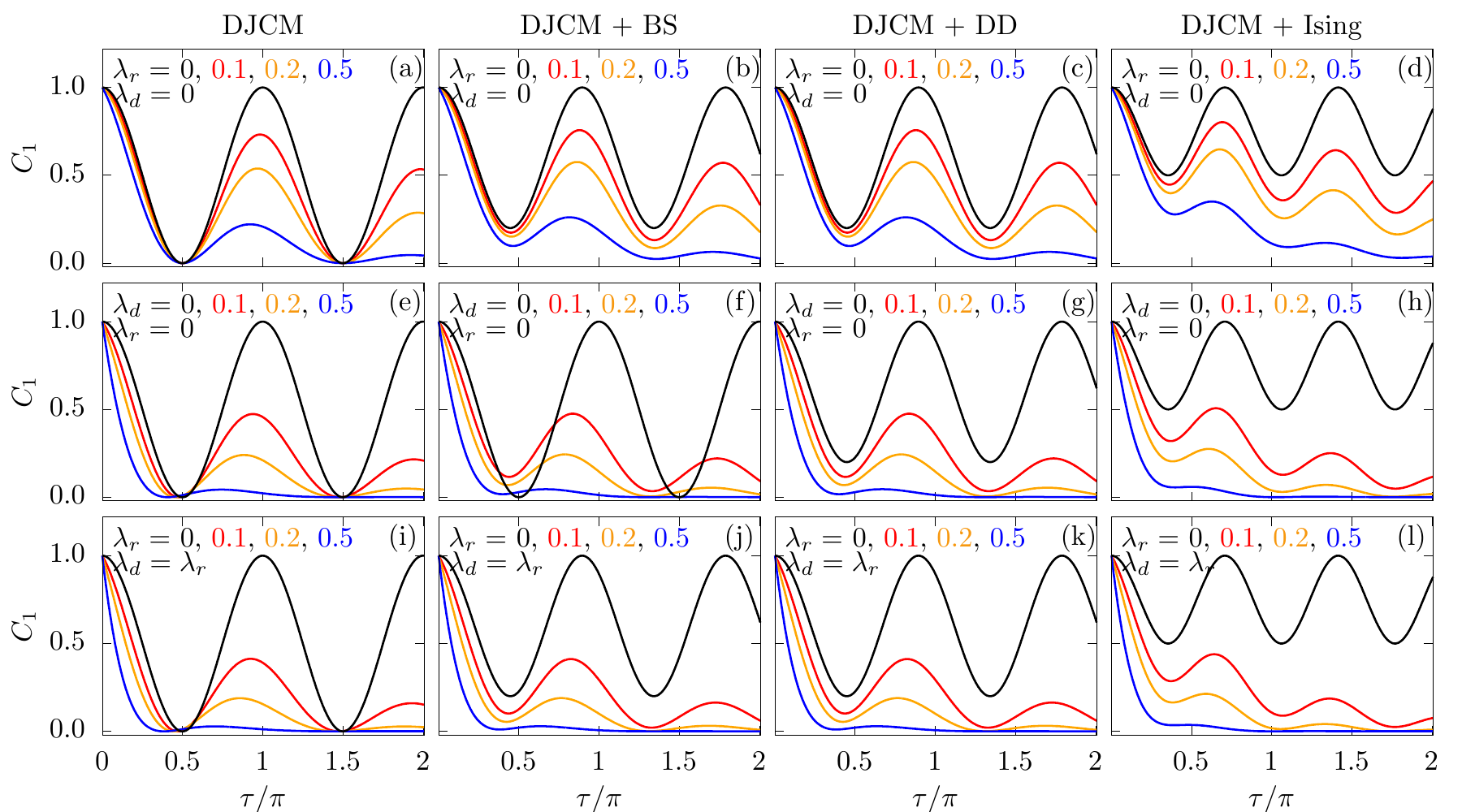}
    \vspace{-1.5ex}
  \caption{Decoherence of the qubit-qubit entanglement through vacuum fluctuations (i.e.,  $\bar{n}_{\text{th}}=0$) are shown for various model systems with different combinations of environmental couplings.  In panels (a-d) we only consider the dissipation channels, i.e,  $\lambda_d=0$. In  panels (e-h) only decoherence through dephasing channels are taken into account ($\lambda_r=0$), and in panels (i-l) both dissipation and decoherence effects are shown for various values of $\lambda_r$ and $\lambda_d$.  Black curves in each plot refer to decoherence free evolution. Initially, qubits are prepared in entangled state $\ket{\psi_{1}(0)}$ with $\phi = \tfrac{\pi}{4}$, while  both oscillators are in  the ground states. $r_{\text{b}} = r_{\text{d}} = r_{\text{i}} = 1$. In all cases we find that the loss of entanglement due to dephasing are much more severe than the  loss of entanglement due to dissipation.}
  \label{fig:decoh1}
\end{figure*}

In Fig.~\ref{fig_stability_nth}, we see the effects of the temperature of the thermal bath  ($\bar{n}_{\text{th}}$) on the stability of entanglement. As dephasing is independent of the the bath temperature, the source of entanglement loss solely comes from $\lambda_r$. We see the manner in which entanglement is suffered  as $\bar{n}_{\text{th}}$ is increased from zero upwards for both the initial states.

\begin{figure*}[ht!]
  \centering
    \includegraphics{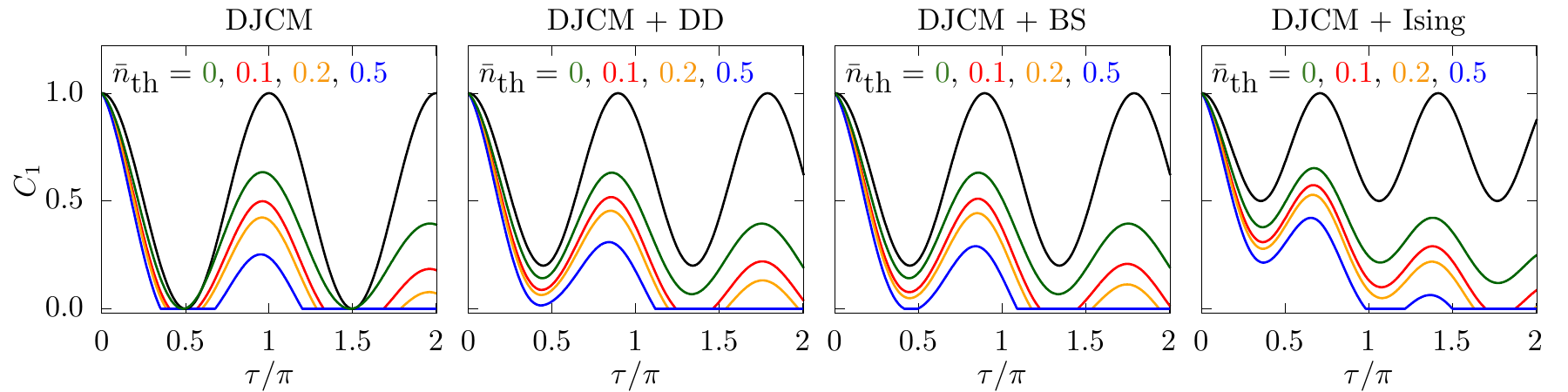}
    \vspace{-1.5ex}
  \caption{Effect of the temperature of the bath ($\bar{n}_\text{th}$) on the qubit-qubit entanglement for various model systems. Black curves in each plot refers to decoherence free evolution while the green curves are for vacuum environment. Initially, qubits are in entangled state $\ket{\psi_{2}(0)}$ with $\phi = \tfrac{\pi}{4}$, while  both oscillators are in  the ground states. $r_{\text{b}} = r_{\text{d}} = r_{\text{i}} = 1$. $\lambda_r=\lambda_d=0.05$.}
 \label{fig:decoh2}
\end{figure*}

With these results at hand, we can now proceed to analyse what happens to the entanglement when we let the entire system interact with the environment.
In Fig.~\ref{fig:decoh1}, we show the losses that arise  for initial qubit-qubit entangled states $\ket{\psi_1(0)}$ for maximal entanglement  $\phi=\tfrac{\pi}{4}$, with both oscillators initially in their respective  ground states. As expected, as the decay rate increases the extent of the loss increases. As the oscillators are assumed to be initially in the ground states, the dissipation of the oscillators and the qubits produce quantitatively similar effects on the dynamics. Also, we see that the effect of  dephasing is much stronger in comparison to dissipation. It is also clear that both the quantitative and qualitative loss of entanglement is independent of the choice of additional interactions. We have verified that these inferences remain valid for other values of $\phi$ as well as for initial state $\ket{\psi_2(0)}$.

Finally, we examine the effects of nonzero $\bar{n}_{\text{th}}$ on the dynamics by keeping the decay rates fixed (Fig.~\ref{fig:decoh2}). Now, all the Lindblad operators (mentioned above) contribute to the dynamics. 
We have further verified that the inferences drawn in this case hold true even if we consider initial coherent state or thermal state oscillators.

\bibliography{references}

\end{document}